\documentclass{article}
\title{Using $ \gamma $-Ray Imaging to Measure Nuclear Lifetimes in the GRETINA Detector}
\author{Dr. Robert Crabbs \\ University of California, Berkeley \\
	\and 
	Dr. I-Yang Lee \\ Lawrence-Berkeley National Laboratory \\
	\and 
	Dr. Kai Vetter \\	Lawrence-Berkeley National Laboratory \\
	}
\date{\today}

\usepackage{amsmath} % Advanced math typesetting
\usepackage{bm} % For typesetting vectors in boldface
\usepackage{graphicx} % Add pictures to your document
\usepackage{pdfpages} % Add PDFs as pages to your document
\usepackage{textcomp} % Includes symbols like the degree-sign
\usepackage{color} % For using different text colors. I'm not shitting you.
\usepackage{multicol,caption} % For making multi-column documents
\usepackage{geometry} % For adjusting margins
\usepackage{hyperref} % For including URL hyperlinks in output

\hypersetup{
  colorlinks   = true, % Colors links instead of ugly boxes
  urlcolor     = blue, % Color for external hyperlinks
  linkcolor    = black, % Color of internal links
  citecolor   = red % Color of citations
}
\newcommand{\scinotation}[2]{#1 $ \times $ 10\textsuperscript{#2}}

\newenvironment{multicolfigure}
  {\par\medskip\noindent\minipage{\linewidth}}
  {\endminipage\par\medskip}

\begin{document}

\maketitle

\begin{abstract}
GRETA, the \textbf{G}amma-\textbf{R}ay \textbf{E}nergy \textbf{T}racking \textbf{A}rray, is an array of highly-segmented HPGe detectors designed to track $ \gamma $-rays emitted in beam-physics experiments. Its high detection efficiency and state-of-the-art position resolution make it well-suited for imaging applications. In this paper, we use simulated imaging data to illustrate how imaging can be applied to nuclear lifetime measurments. This approach can offer multiple benefits over traditional lifetime techniques such as RDM. 
\end{abstract}

\begin{multicols}{2}

%%%%%%%%%%%%%%%%%%%%%%%%%%%%%%%%%%%%%%%%
% Section: Introduction
%%%%%%%%%%%%%%%%%%%%%%%%%%%%%%%%%%%%%%%%

\section{Introduction}
\label{sec:introduction}

Gamma-ray tracking \cite{gamma_ray_tracking_detectors} \cite{new_concepts_in_gamma_detection} is a major advance in gamma-ray spectroscopy. A 4$ \pi $ tracking-array would be a powerful instrument for a broad range of experiments in low-energy nuclear science \cite{nsac_long_range_plan} \cite{nupecc_long_range_plan}, especially for the nuclei far from the line of stability. Developments of these instruments are underway \cite{agata_and_greta} both in the US (GRETINA/GRETA) \cite{gretina_performance_whitepaper} \cite{gretina_spectroscopy_performance} \cite{greta_website} and Europe (AGATA) \cite{agata_whitepaper} \cite{agata_website}. The GRETA collaboration has built a partial array, called the Gamma-Ray Energy Tracking IN-beam Array (GRETINA). The array's primary purpose is for high-efficiency, high-precision $ \gamma $-ray spectroscopy. Its position sensitivity allows the sequencing of photon tracks via Compton kinematics, which is used, among other things, to estimate the emission angle of a gamma ray relative to its parent nucleus’s velocity. This is key for spectroscopy because it allows us to correct Doppler shifts in the lab-frame photon energy.
\par
GRETINA's excellent position- and energy-resolution also open the door for gamma-ray imaging. \cite{gamma_ray_tracking_opportunities} \cite{compton_imaging_in_gretina} \cite{doppler_imaging_in_gretina} If we can accurately reconstruct the position distributions of photon sources from in-beam experiments, we can measure nuclear lifetimes more efficiently than is possible with current methods such as the Recoil-Distance Method (RDM). \cite{lifetime_measurement_74rb} \cite{rdm_triplex_plunger} Where RDM generally requires multiple measurements to find a lifetime, imaging can theoretically deliver full de-excitation curves from a single run. Lifetimes can be extracted directly by fitting to these curves.
\par
In previous research, we investigated performance of two gamma-ray imaging techniques with the GRETINA array. This paper shows how one might apply our imaging results to lifetime measurements, including methods to correct for imperfect imaging resolution. If the detector's imaging response can be characterized well enough, even very poor imaging resolution can still be used to measure nuclear lifetimes.

%%%%%%%%%%%%%%%%%%%%%%%%%%%%%%
% Section: Intro to Lifetime Measurements
%%%%%%%%%%%%%%%%%%%%%%%%%%%%%%
\section{Introduction to Lifetime Measurements}
\label{sec:intro_to_lifetime_measurements}

The simplest lifetime measurement involves a parent nucleus with proper lifetime $ \tau_0 $ and lab-frame velocity $ \beta $, which emits a single characteristic photon of energy $ E_0 $ in the CM-frame. The gamma de-excitation of these parent nuclei has an exponential time-dependence, which produces an exponential source distribution along the beamline. (Note that a nucleus's travel distance directly corresponds to its emission time.) Our job is to gather enough statistics to produce an image of this exponential decay, and then extract the lifetime from it. This process can be made independent of the choice of imaging mode -- what matters is the imaging response and effective imaging resolution. Images from known calibration measurements or simulations can be used to characterize this resolution.
\par
RDM is one common technique for lifetime measurements. A ``degrader'' foil is placed downstream of the target, which slows the parent nuclei to a lower speed. Photons emitted upstream of this degrader will exhibit a Doppler-shift in the lab frame, while photons emitted downstream will have little to no shift. One can therefore obtain the fraction of nuclei that de-excite upstream of the degrader by measuring counts in an energy spectrum. By moving the degrader to several positions and repeating the measurement, one obtains the fraction of nuclei that de-excite as a function of distance (i.e. flight time). Fitting an exponential decay function to these samples then yields the proper lifetime, $ \tau_0 $.  
\par
With gamma-ray imaging, we trace each photon back to its emission point along the beamline, then convert each distance to a time-of-flight. With enough counts, we can build a de-excitation curve and extract the lifetime. The appeal of this technique is that it can reconstruct an entire de-excitation curve with a single measurement, without additional downstream equipment. It is well-suited to timescales of 10's of picoseconds to a few nanseconds.
\par 
The two imaging methods presented here require knowledge of the photon interaction sequence in the detector. A $ \gamma $-ray typically interacts several times in an HPGe detector before being fully-absorbed, resulting in a sequence of hits in the detector denoted $ \bm{X_1} $ to $ \bm{X_N} $. (Figure \ref{fig:compton_imaging_geometry}) However, we cannot directly measure the interaction sequence because the detector electronics are not fast enough to resolve the differences in timing. Instead, we use Compton sequencing to deduce the sequence. \cite{compton_sequencing_in_gretina}
\par
In Compton imaging, we can define a ``Compton cone'' from detected energy depositions and the locations of the first 2 interactions. (Figure \ref{fig:compton_imaging_geometry}) Each cone shows the possible directions from which a photon came as it entered the detector, and is uniquely defined by its vertex $ \bm{X_C} $, central axis $ \bm{\hat{V}_C} $, and cosine of opening angle $ \mu_C = \cos{\theta_C} $.
\par
The intersections of these Compton cones with the recoil beam are the possible emission points of the photons. Note that a cone will intersect a beam at (up to) 2 points. In many cases, one intersection will be \textit{upstream} of the beam target -- clearly an unphysical solution. Throwing out the bad intersection yields a single unambiguous emission point on the beamline. In other cases, both intersections are downstream of the target, resulting in ambiguity. Due to imperfect detector resolution, it's even possible for the cone not to intersect the beamline at all. These and other details are covered in Reference \cite{compton_imaging_in_gretina}. 

\begin{multicolfigure}
\centering
\includegraphics[width=1.0\textwidth]{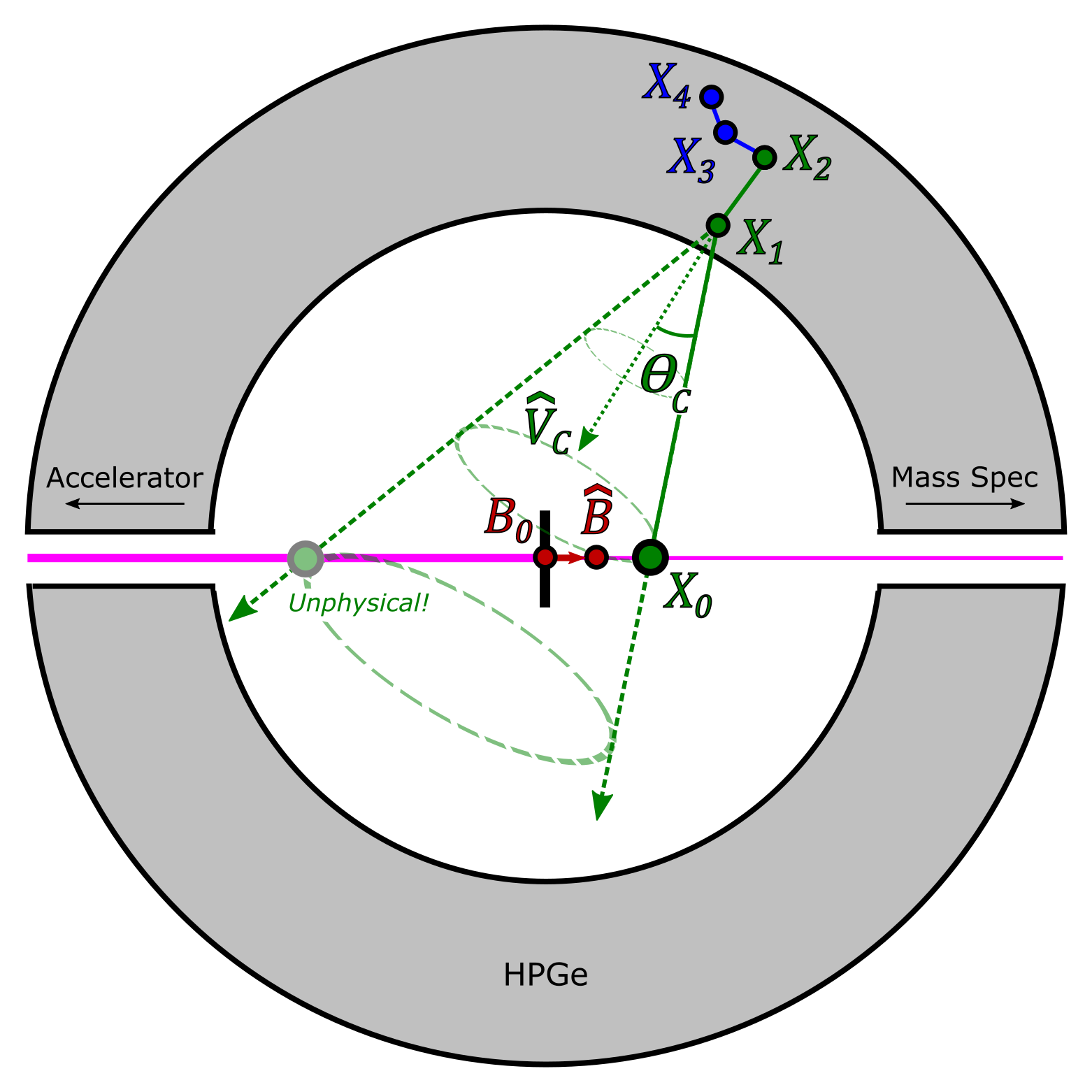}
\captionof{figure}{Geometry used in Compton imaging}
\label{fig:compton_imaging_geometry}
\end{multicolfigure}

\begin{multicolfigure}
\centering
\includegraphics[width=1.0\textwidth]{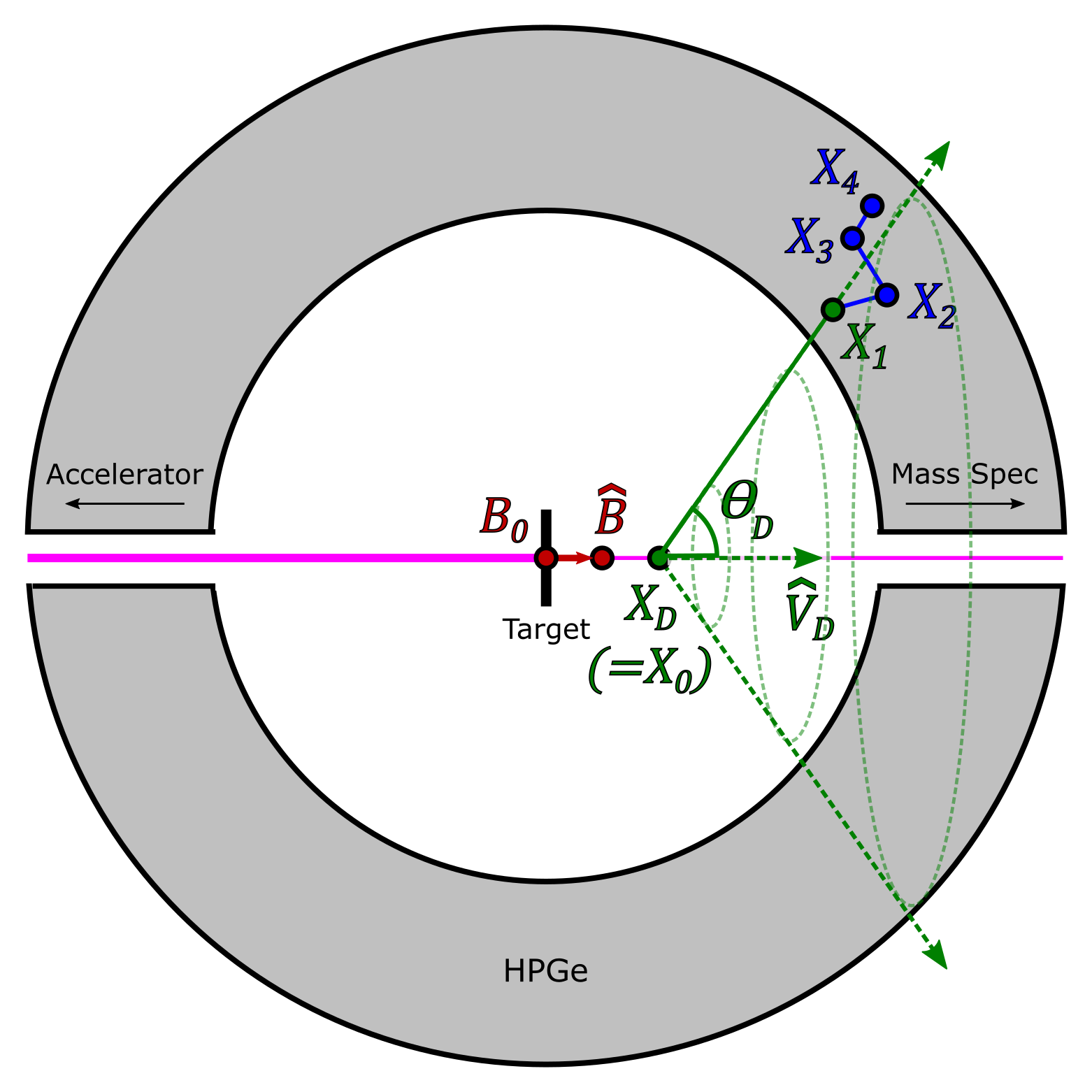}
\captionof{figure}{Geometry used in Doppler-shift imaging}
\label{fig:doppler_imaging_geometry}
\end{multicolfigure}

\par
In Doppler-shift imaging, we use the observed Doppler-shift in photon energy to locate the emission point $ \bm{X_0} $. If we know the photon parent's lab-frame velocity and the emitted $ \gamma $-ray's characteristic rest frame energy $ E_{CM} $, we can use the detected lab-frame energy $ E_{Lab} $ to determine the emission angle:
\begin{align}
\label{eq:doppler_shift_emission_angle}
\cos{\theta} = \frac{1}{\beta} \left(1 - \frac{E_{CM}}{\gamma E_{Lab}} \right)
\end{align}
where $ \beta c $ is the parent's speed in the lab-frame, $ \gamma $ is its corresponding Lorentz factor, and $ \theta $ is the lab-frame emission angle.
\par
From this angle we can draw a ``Doppler cone'' representing all emission vectors that would result in the observed Doppler-shift. (Figure \ref{fig:doppler_imaging_geometry}) The cone's vertex $ \bm{X_D} $ is located at the photon emission point and its central axis $ \bm{\hat{V}_D} $ is directed along the beamline. Its opening angle $ \theta_D $ is given by Equation \ref{eq:doppler_shift_emission_angle}.
\par
Note that the cone must pass through the location of the first interaction in the detector ($ \bm{X_1} $). This provides a measured anchor point that we can use to deduce the emission point, $ \bm{X_0} $. Reference \cite{doppler_imaging_in_gretina} discusses Doppler-shift imaging in much greater detail.
\par
To understand the images produced by Compton or Doppler-shift imaging, we need to look at the underlying structure of the de-excitation curves. Consider the simplest case where a single nuclear transition neither feeds -- nor is fed by -- any other transitions. In this situation, there is a single lifetime $ \tau_0 $ to consider, and the density of emissions downstream of the target will follow an exponential decay:
\begin{align}
\label{eq:exponential_decay}
S(z) &= A e^{-\mu z} \\
&= A e^{-z / \gamma \beta c \tau_0}
\end{align}
where $ z $ is the distance from the target, $ \mu = 1/\gamma \beta c \tau_0 $, and A is a scaling factor. This factor is found by normalizing the curve to the number of emissions $ N_{counts} $ observed between the target and end of imaging window ($ z_{end} $):
\begin{align}
N_{counts} &= \int_0^{z_{end}} Ae^{-\mu z} dz = (A / \mu) (1 - e^{-\mu z_{end}}) \\
A &= N_{counts} \frac{\mu}{1 - e^{-\mu z_{end}}}
\end{align}
Note that A has the correct units for a density function (counts per length).
\par
When the state of interest is part of a de-excitation chain, the shape of the curve no longer follows a simple exponential. For example, consider a two-state system where the parent state has lifetime $ \tau_{0,p} = 1 / \gamma \beta c \mu_p $ and the daughter state has lifetime $ \tau_{0,d} = 1 / \gamma \beta c \mu_d $. The source distribution $ S_d(z) $ is found:
\begin{align}
S_p(z) &= A_p e^{-\mu_p z} \\
S_d(z) &= A_d (e^{-\mu_p z} - e^{-\mu_d z}) \\
A_d &= A_p \left( \frac{\mu_p}{\mu_d - \mu_p} \right) \\
&= N_{counts} \left( \frac{\mu_p}{1 - e^{-\mu z_{end}}} \right) \left( \frac{\mu_p}{\mu_d - \mu_p} \right)
\end{align}
The shape of the de-excitation curve is determined by the the relative lifetimes of the parent and daughter. For example, in a system where the parent is very long-lived compared to the daughter, such that $ \mu_d >> \mu_p $:
\begin{equation}
S_d(z) = N_{counts} \left( \frac{\mu_p}{1 - e^{-\mu z_{end}}} \right) \left( \frac{\mu_p}{\mu_d} \right) (1 - e^{-\mu_d z})
\end{equation}
This leads to secular equilibrium, where the countrate from the daughter is limited by the de-excitations of the parent. If the parent lifetime is too long, then we will not be able to use imaging to study the lifetime of either state. The source distributions will appear flat across the entire imaging range.
\par
The source distributions of more complex de-excitation chains can be solved using the Bateman equations. In such cases, we must have prior knowledge of the lifetimes for all parent states feeding the state of interest. Complex chains may also produce significant background -- Doppler-shifts can cause peaks to overlap even if they are several hundred keV apart in the CM-frame. For this study, we will focus on simple single-transition systems.

%%%%%%%%%%%%%%%%%%%%
% Section: Curve Fitting and the Inverse Imaging Problem
%%%%%%%%%%%%%%%%%%%%
\section{Curve Fitting and the Inverse Imaging Problem}
\label{sec:curve_fitting_and_imaging}

In lifetime measurements, we seek to reproduce the source distribution for nuclei that de-excite downstream of the target. The ideal distribution is a simple exponential decay function given by Equation \ref{eq:exponential_decay}. However, due to finite detector performance and other real-world considerations, imaging does not recreate source distributions with perfect accuracy. \cite{position_sensitivity_in_hpge} \cite{greta_detector_performance} This is illustrated in Figure \ref{fig:true_vs_imaged_source_distributions}, which compares the true underlying source distribution and the corresponding \textit{imaged} emission points from a Compton imaging test. (Similar images could be produced for Doppler-shift imaging as well.) For any true emission point $ z_0 $, there is a distribution $ P(z,z_0) $ that reflects the probability that the emission will instead register at $ z $. We call this probability density function the ``detector response''. For the rest of this discussion, we will divorce ourselves from the specific type of imaging method used, and focus instead on the imaging \textit{resolution} as the primary parameter of interest.

\begin{multicolfigure}
\centering
\includegraphics[width=1.0\textwidth]{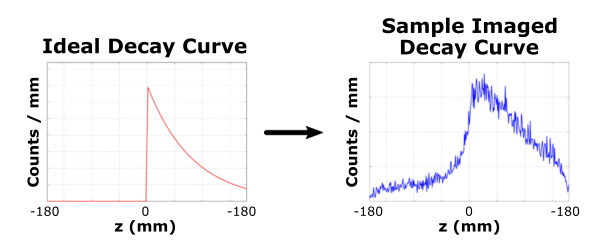}
\captionof{figure}{True vs. imaged source distributions}
\label{fig:true_vs_imaged_source_distributions}
$ \tau_0 $ = 859 ps, $ \beta $ = 0.3 \\
$ \sigma_{xyz} $ = 3.0 mm, $ \sigma_E $ = 2.0 keV \\
\end{multicolfigure}

\par
Imperfect detector response will produce an imperfect imaged representation of a true source distribution. By discretizing $ z $ and $ z_0 $ into bins, this convolution can be represented by the matrix equation:
\begin{equation}
\label{eq:source_to_image_convolution}
R(z) = P(z,z_0) ~ S(z_0)
\end{equation}
where $ R(z) $ gives the imaged source distribution. Note that $ R(z) $ is an $ M \times 1 $ column vector, where M is the number of distinct bins forming the imaging space on the beam axis. $ S(z_0) $ is an $ N \times 1 $ column vector, where N is the number of distinct $ z $'s for which a detector response distribution is measured. This makes $ P(z,z_0) $ an $ M \times N $ matrix, where the \textit{nth} column represents the detector response function for the \textit{nth} $ z_0 $ bin.
\par
Because imaged de-excitation curves may deviate significantly from an exponential shape, we need to account for imaging response in the curve-fitting process used to extract lifetime. The convolution in Equation \ref{eq:source_to_image_convolution} will tend to move counts from the peak of the exponential to somewhere upstream of the beam target, rounding the top of the imaged source distribution and making the slope of the decay curve more shallow. Therefore, fitting a simple exponential to this shape will return a measured lifetime longer than the true lifetime.
\par
Geometric efficiency can also play a role. GRETINA is not a full $ 4\pi $ array, and so photon emissions can escape the central cavity without passing through a detector. GRETINA's geometric efficiency is 25\% for an isotropic point source at the center of the array, but things are more complicated for moving sources. Emissions become more forwardly-directed in the lab frame as the source velocity increases. In addition, the detector solid angle ``seen'' by a parent nucleus changes as the nucleus moves down the beamline. Geometric efficiency is therefore a function of both source velocity and emission location. Because of this, the measured source distribution will not be perfectly exponential -- even with perfect imaging resolution. 
\par
The effect is shown in Figure \ref{fig:geometric_efficiency_geant4}. We used Geant4 to simulate photon emissions along the beamline, and then determined which ones hit a GRETINA detector module (regardless of complete energy deposition). The emissions were grouped into three categories -- total emissions, \textit{detected} emissions, and emissions which missed the detectors entirely. The thicker lines are idealized de-excitation curves for each category, given the number of emissions and the known parent lifetime and velocity. One can see that the detected emissions fall short of the exponential shape due to losses in geometric efficiency. In general, each detector response function $ S(z_0) $ should be renormalized to the geometric efficiency at $ z_0 $.

\begin{multicolfigure}
\centering
\includegraphics[width=1.0\textwidth]{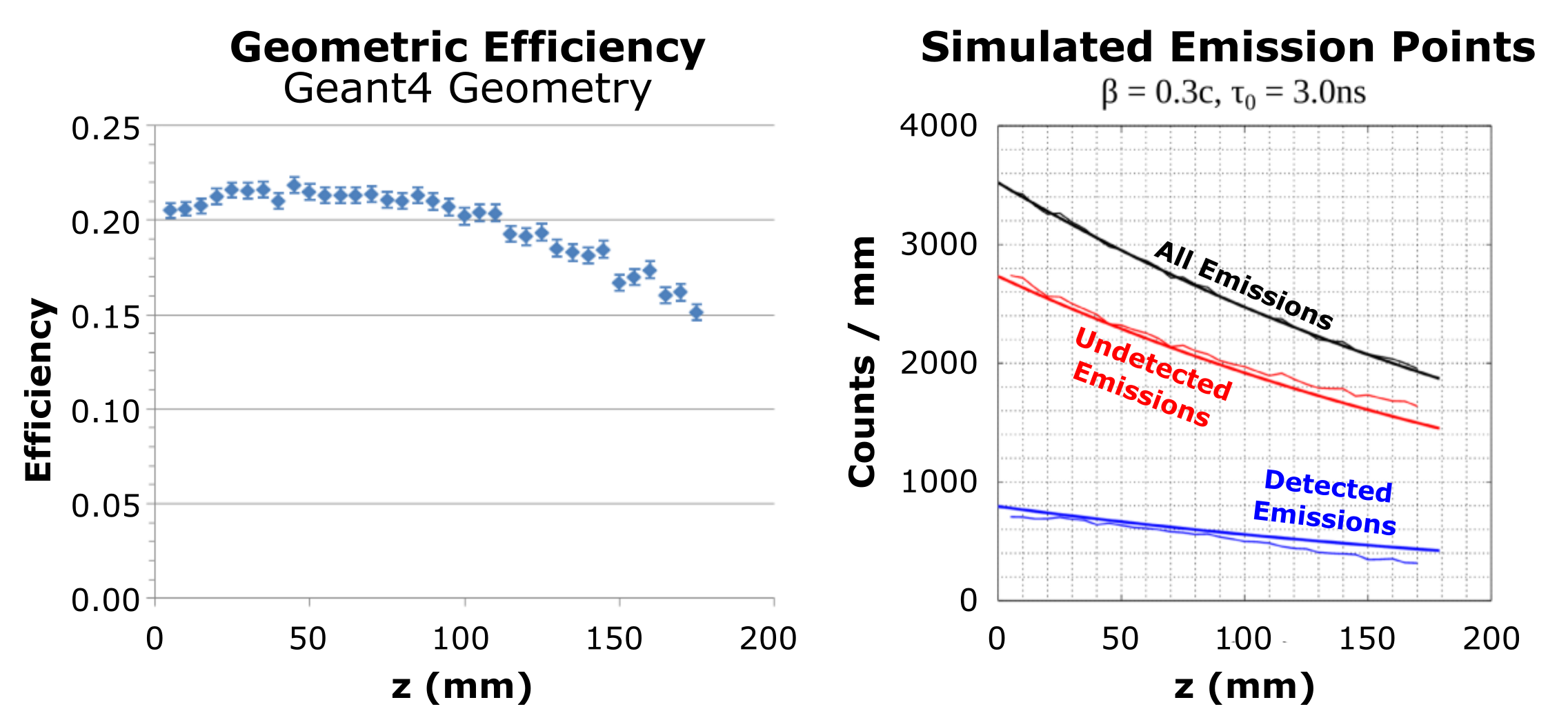}
\captionof{figure}{Geometric efficiency in GRETINA}
\label{fig:geometric_efficiency_geant4}
(Left): Overall efficiency along the beamline \\
(Right): Deviations caused in ideal exponentials
~\\
\end{multicolfigure}

\par
There are common methods to correct for non-idealities in a source distribution. The imaging response $ P(z,z_0) $ is a measurable property of the specific detector and imaging method employed. One might obtain it through a calibration experiment using a nucleus of known lifetime. For this study we calculated simulated detector response curves from a Geant4 model of the GRETINA detector. \cite{UCGretina_model} With a $ P(z,z_0) $ matrix in hand, one can then use numerical methods such as Singular-Value-Decomposition (SVD) or $ \chi^2 $ Minimization to account for the imaging response. \cite{chi2_fitting_method} \cite{SVD_fitting_method} 
\par
In SVD, the goal is to find the pseudo-inverse of the detector response, $ P^{-1}(z_0,z) $, in order to calculate the ``true'' source distribution:
\begin{equation}
S(z_0) = P^{-1}(z_0,z) R(z)
\end{equation}
Theoretically, this correction should let us fit an exponential directly to the curve $ S(z_0) $. However, noise amplification is a well-known problem in SVD. \cite{noise_in_SVD} The algorithm is also sensitive to how $ z $ and $ z_0 $ bins are discretized, which introduces an unwelcome additional parameter to tune. \cite{ideal_binning}
\par
The $ \chi^2 $ Minimization method avoids these issues, though it is more complicated and computationally-intensive. In SVD, we work backwards from the image $ R(z) $ to calculate an approximate source distribution $ S(z_0) $. The $ \chi^2 $ approach is the reverse -- we start with a \textit{hypothetical} source $ S'(z_0) $, then convolve it with the known response matrix $ P(z,z_0) $ to arrive at a hypothetical image, $ R'(z) $. (See Figure \ref{fig:SVD_vs_chi2_minimization}) We then compare this hypothetical $ R'(z) $ image with the measured image, $ R(z) $, computing a $ \chi^2 $ value:
\begin{align}
\label{eq:chi2_lifetime_measurement}
\chi^2 &= \sum_{n=1}^{N} \frac{(R(z_{n}) - R'(z_{n}))^2}{\sigma_{R'(z_{n})}^2} \\
&= \sum_{n=1}^{N} \frac{(R(z_{n}) - R'(z_{n}))^2}{R'(z_{n})}
\end{align}
Here, $ \sigma_{R'(z_{n})}^2 = R'(z_{n}) $, and N is the number of $ z $ bins used to characterize image distributions. (The first \& last ``catch-all'' bins are omitted to avoid edge effects in the images.) This process is repeated for many different hypothetical source distributions, and the one with the smallest $ \chi^2 $ value is selected as the best fit (hence ``$ \chi^2 $ Minimization'').

\begin{multicolfigure}
\centering
\includegraphics[width=1.0\textwidth]{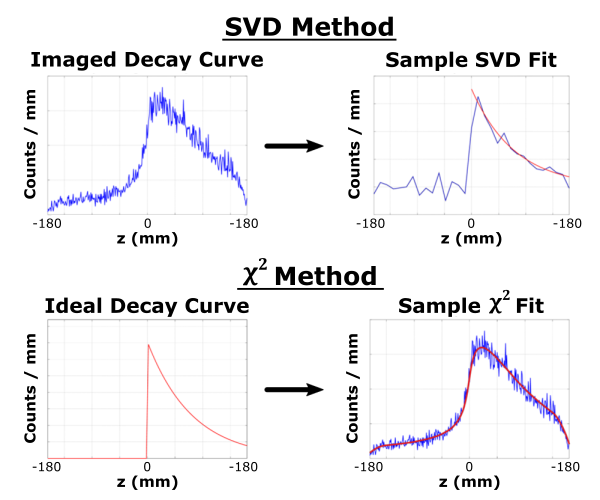}
\captionof{figure}{SVD vs. $ \chi^2 $ Minimization curve-fitting}
\label{fig:SVD_vs_chi2_minimization}
~\\
(Top): SVD transforms an image back into an ideal exponential curve \\
(Bottom): $ \chi^2 $-Minimixation transforms an ideal curve into a best-fit \textit{image}
\\~
\end{multicolfigure}

\par
For simple single-transition systems, a source distribution $ S(z_0) $ can be fully parameterized by the nuclear lifetime, $ \tau_0 $. We can therefore iterate on the value of $ \tau_0 $ to generate progressively better fits to the observed image $ R(z) $. The first iteration starts with a guess at the range of possible parent lifetimes (for example, 250 ps to 4 ns). In this study, we break this range into 20 evenly-spaced $ \tau_0 $ values. A $ \chi^2 $ value is computed for each. We then construct a new range of 20 evenly-spaced points between the left- and right-hand neighbors of the $ \tau_0 $ with the minimum $ \chi^2 $ value. This is repeated until the fitted lifetime converges to within 0.1 ps, which generally occurs after 5-6 iterations. 
\par
At the end of this process, we have $ \chi^2 $ as a function of hypothetical values of $ \tau_0 $. This curve enables us to estimate the uncertainty in the best-fit lifetime from a single measurement. To do so, we find the values of $ \tau_0 $ for which $ \chi^2 = \chi_{min}^2 + 1 $. These represent the $ \pm1\sigma $ error bars for the lifetime. 

\begin{multicolfigure}
\centering
\includegraphics[width=1.0\textwidth]{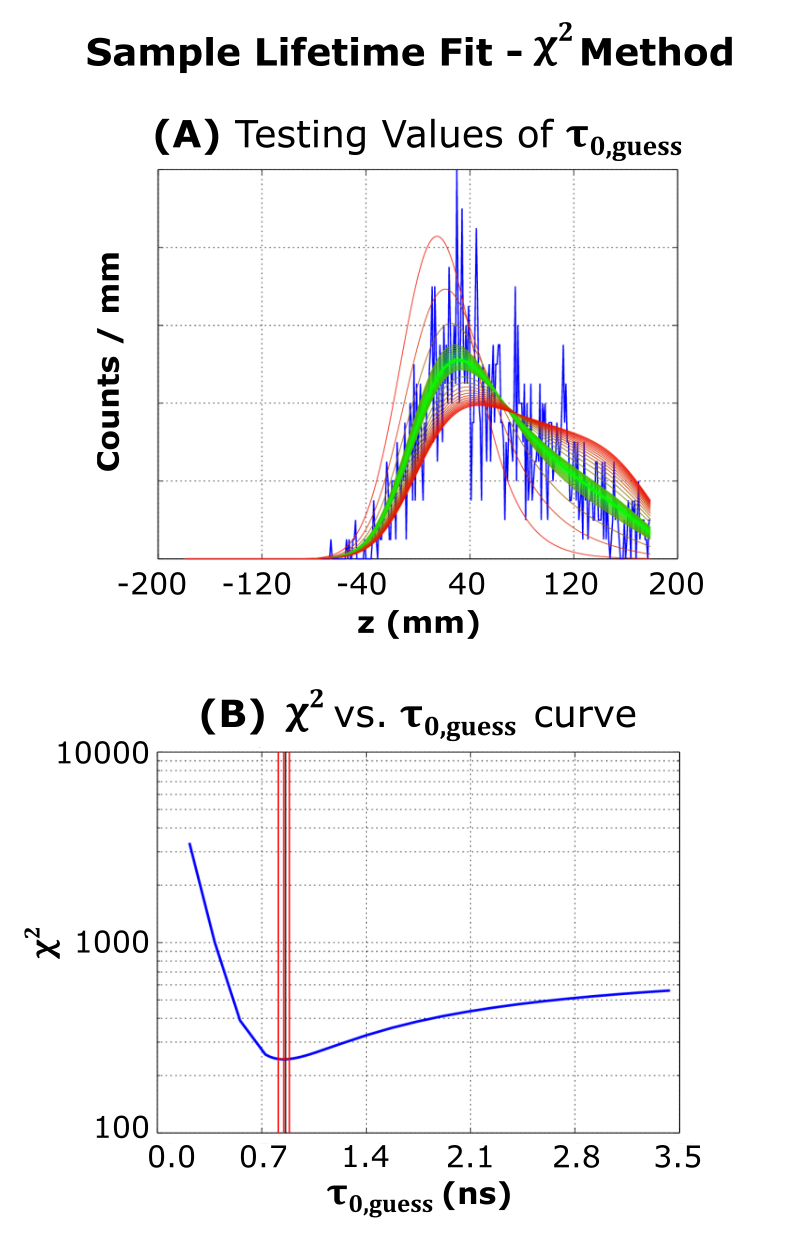}
\captionof{figure}{A sample $ \chi^2 $ Minimization fit}
\label{fig:sample_lifetime_fit}
\end{multicolfigure}

\par
Figures \ref{fig:sample_lifetime_fit}A-B show a sample $ \chi^2 $ fit. In Figure \ref{fig:sample_lifetime_fit}A, the hypothesized images (red and green) are compared to the raw data (blue). The green lines correspond to lifetimes close to the true value of $ \tau_0 $, and have the smallest $ \chi^2 $ values. By contrast, the red lines correspond to lifetimes that are either too short or too long, and which therefore produce generally poor fits to the measured de-excitation curve. In other words, the green lines converge to our best guess for the true value of $ \tau_0 $. Figure \ref{fig:sample_lifetime_fit}B shows the $ \chi^2 $ vs. $ \tau_{0,guess} $ curve. Here, the red lines show the limits of the $ \chi^2 + 1 $ estimate on the uncertainty.

%%%%%%%%%%%%%%%%%%%%
% Section: Simulated Lifetime Measurements
%%%%%%%%%%%%%%%%%%%%
\section{Simulated Lifetime Measurements}
\label{sec:simulated_lifetime_measurements}

Three of the primary factors in curve-fitting are the imaging resolution, $ \sigma_{img} $; the number of imaged emissions, $ N_{img} $; and the decay parameter, $ \mu = 1 / \gamma \beta c \tau_0 $, a combination of the beam velocity and proper lifetime. Each of these factors affects the precision that we can achieve with lifetime measurements using gamma-ray imaging. We used a series of Monte Carlo simulations to quantify this precision, generating hypothetical source images for each combination of $ \sigma_{img} $, $ N_{img} $, and $ \mu $. 
\par
Generating hypothetical source images is fairly straightforward. First, we define values for $ \tau_0 $ and $ \beta $ to formulate an ideal de-excitation curve. This allows sampling the de-excitation curve for a true emission point, $ z_0 $. Next, we sample the detector response distribution at $ z_0 $ to find the \textit{imaged} emission point, $ z $. This process is repeated until the desired number of counts $ N_{samples} $ is reached. With a simulated imaged de-excitation curve, we can use the curve-fitting described in Section \ref{sec:curve_fitting_and_imaging} to extract a lifetime. Lastly, to estimate precision of the calculated lifetime, we run this process for a large number of simulated de-excitation curves (e.x. $ N_{runs} $ = 1000). 
\par
For this study, we assumed simple Gaussian detector response curves with the desired imaging resolution, $ \sigma_{img} $. As can be seen from Figure \ref{fig:true_vs_imaged_source_distributions}, however, the imaging responses can deviate from this idealized shape. Compton imaging responses have long tails to either end of the imaging window (and the peaks sit atop wide backgrounds). Doppler-shift imaging responses have much more Gaussian-shaped peaks, but also have a long, low tail downstream of the true emission point.
\par
Repeating a set of $ N_{runs} $ simulations produces a distribution of \textit{calculated} lifetimes for a given imaging resolution, number of counts, and underlying lifetime. While the standard deviation of this distribution gives the precision obtainable in a lifetime measurement, this approach would require running the same experiment $ N_{runs} $ times. Fortunately, we can also estimate uncertainty from a \textit{single} run using the $ \chi_{min}^2 + 1 $ method described in curve fitting. The single-run estimates for $ \sigma_{\tau} $ are largely in agreement with the standard deviations found from multi-simulation tests. Figure \ref{fig:estimating_precision} illustrates this -- the single- and multi-measurement $ \sigma_\tau $ values are within a 10\% margin (less than the standard deviation of the single-measurement estimates). This therefore provides confidence in using the $ \chi_{min}^2 + 1 $ method in real-world experiments, where we do not have the luxury of running 1000 trials of the same test.

\begin{multicolfigure}
\centering
\includegraphics[width=1.0\textwidth]{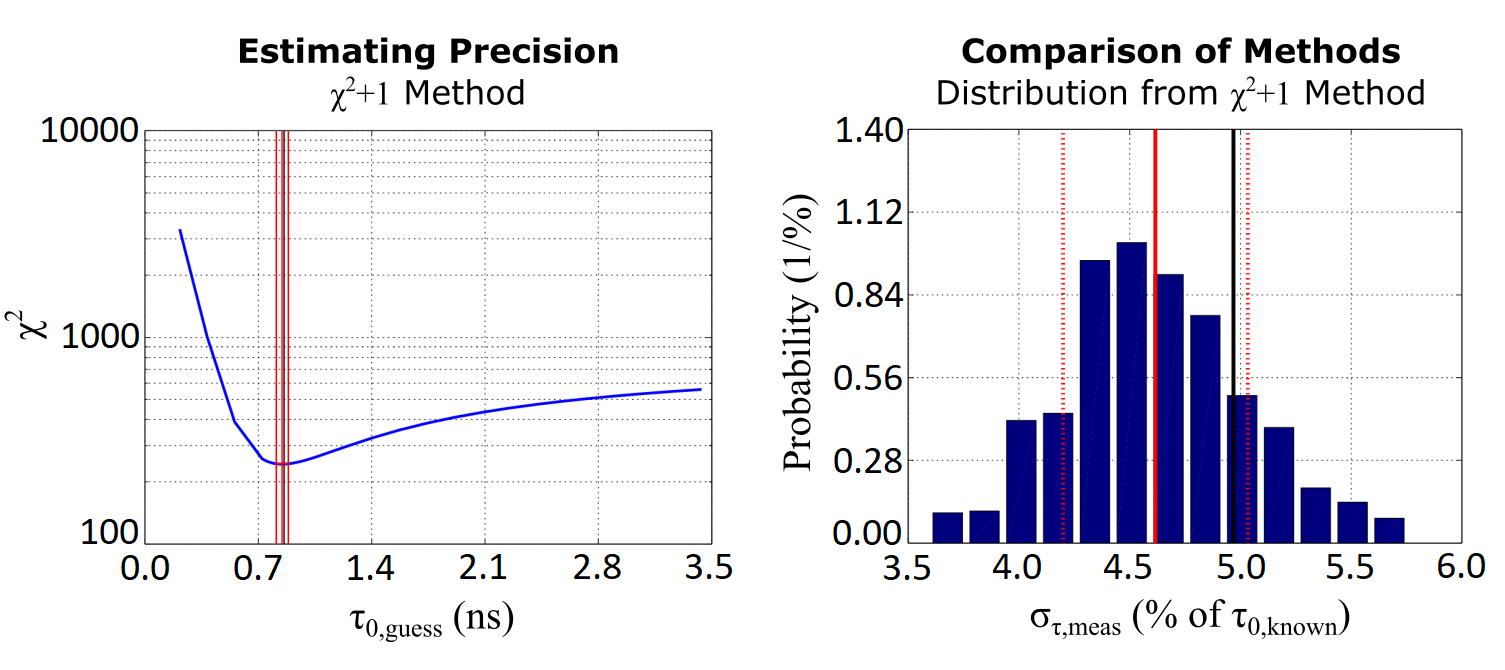}
\captionof{figure}{Estimating precision in lifetime}
\label{fig:estimating_precision}
\end{multicolfigure}

\par
Gathering enough counts is a major challenge in lifetime measurements. The nuclear state of interest may have a \textit{very} small production cross-section. On top of that, not every nucleus produced in the target will necessarily emit a photon within the imaging range. (The fraction depends on the beam velocity and the nuclear lifetime.) The detector also has less-than-perfect geometric and intrinsic efficiencies, and therefore will not detect every photon emission. GRETINA covers about $ 1\pi $ steradians as seen from the center of its inner cavity, and has a peak-to-total ratio of approximately 60\% for \textsuperscript{137}Cs. The placement of its detector modules is also an important consideration -- because the parent nuclei are moving relativistically, the lab-frame emissions will tend to be forward-directed. The exact angular distribution depends on the beam velocity and the angular distribution in the CM-frame. 
\par
Gamma-ray imaging methods also have imperfect imaging efficiency. Not every track with full energy deposition yields an emission point. In Compton imaging, tracks are rejected for giving ambiguous emission points or emission points outside the imaging range (-180 mm $ < z < $ 180 mm). They can also be rejected for having small Compton lever arms or large sequencing FoMs. In Doppler-shift imaging, tracks are rejected if they yield emission points outside the imaging range. When no data filters are applied, typical imaging efficiency for 1.0 MeV CM-frame emissions at 0.3c is 30\% for Compton imaging and 72\% for Doppler-shift imaging. These numbers drop to 5-10\% for Compton imaging and 10-20\% for Doppler-shift imaging depending on the choice of data cut(s). \cite{compton_imaging_in_gretina} \cite{doppler_imaging_in_gretina}
\par
To illustrate, consider an experiment with beam intensity $ I $, production cross-section $ \sigma $, atom density $ n $, target thickness $ x $, geometric efficiency $ \epsilon_{g} $, intrinsic efficiency of $ \epsilon_{i} $, and imaging efficiency $ \epsilon_{m} $. Suppose the parent nuclei have velocity $ \beta c $ and proper lifetime $ \tau_0 $, with a branching ratio $ B $ for the transition of interest. If the beam runs for time $ t $, then the total number of nuclei produced is:
\begin{equation}
N_{tot} = t I B (1 - e^{-\Sigma x}) \simeq t I B \Sigma x 
\end{equation}
where $ \Sigma = n \sigma $. The total number of emissions created within the imaging window depends on the target position. Suppose the target is placed at $ z = $ (0.0 mm, 0.0 mm, 0.0 mm) and the right edge of the imaging window is at $ z_r $. Then the number of emissions that \textit{could} be detected is given by:
\begin{align}
N_{emit} &= N_{tot} ( 1 - e^{-\mu z_r}) \\
\mu &= 1 / \gamma \beta c \tau_0
\end{align}
Only a subset of these can be used for imaging ($ N_{samples} $). Constructing a de-excitation curve therefore requires a total beamtime of $ t_N $:
\begin{align}
N_{samples} &= N_{emit} (\epsilon_{g} \epsilon_{i} \epsilon_{m})  \\
&= (t I B \Sigma x) ( 1 - e^{-\mu z_r}) (\epsilon_{g} \epsilon_{i} \epsilon_{m}) \\
t_N &= \frac{N_{img}}{(I B \Sigma x) ( 1 - e^{-\mu z_r}) (\epsilon_{g} \epsilon_{i} \epsilon_{m})}
\end{align}

For example, consider a typical secondary beam (\scinotation{1.5}{4} particles/s) with a reaction of cross-section of $ \sigma $ = 10.0 mb. Suppose the target is $ x $ = 1.0 mm thick with density $ n $ = \scinotation{1.50}{22} atoms/cm\textsuperscript{3}, and let $ B = 0.1 $, $ \beta $ = 0.30, $ \tau_0 $ = 1.0 ns, $ z_r $ = 180 mm, $ \epsilon_{g} $ = 0.25, $ \epsilon_{i} $ = 0.60, and $ \epsilon_{m} $ = 0.20. All told, gathering 100 usable reconstructions requires 48 hrs of beamtime. 
\par
\begin{figure}[ht]
\centering
\includegraphics[width=1.0\textwidth]{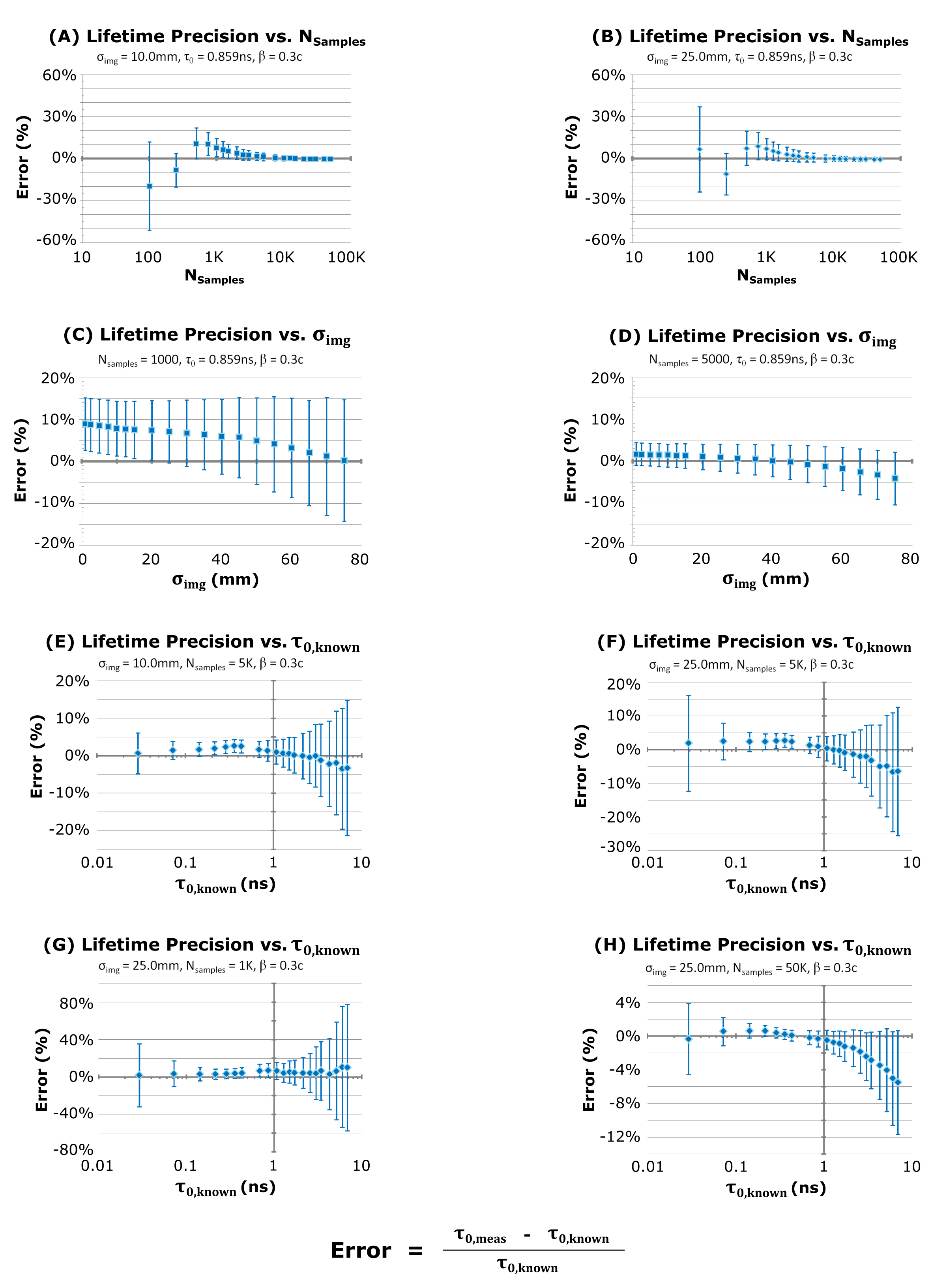}
\caption{Effects of several experimental variables on lifetime precision}
\label{fig:precision_vs_experimental_variables}
\end{figure}

Table 1 shows how overall precision in lifetime depends on counting statistics and imaging resolution. For this set of simulations, parent CM-frame lifetime and lab-frame velocity were kept constant at 0.859 ns and 0.3c, respectively. The number of reconstructions in each de-excitation curve ($ N_{samples} $) ranged from 100 to 50K samples while the detector's Gaussian imaging resolution ($ \sigma_{img} $) ranged from 1.0 mm to 75.0 mm, resulting in a total of 399 distinct runs. Figures \ref{fig:precision_vs_experimental_variables}A-B shows the dependence on $ N_{samples} $ for two imaging resolutions -- 10.0 mm and 25.0 mm. These values reflect typical imaging resolution of Doppler-Shift and Compton imaging, respectively.  
\par
Figures \ref{fig:precision_vs_experimental_variables}C-D illustrates the dependence on $ \sigma_{img} $ for two distinct values of $ N_{samples} $. There is some fluctuation in the measured value of $ \tau_0 $ in the low-count regime ($ N_{samples} < $ 1000), as well as an apparent systematic error in the average measured value of $ \tau_0 $ (which manifests as a monotonic convergence as imaging resolution improves). This effect is attributed to the random number generation used in the simulations. We use the same random number seed to start each set of lifetime simulations to preserve reproducibility. Using a different seed between runs removes the monotonic trend, although the measured lifetime still converges as expected to the nominal value (0.859 ns).
\par
The systematic error decreases to generally less than 2\% with $ N_{samples} $ = 5000. Meanwhile, the precision $ \sigma_\tau $ steadily worsens with increasing width of detector imaging resolution, as expected. The dependence is surprisingly weak, however -- even with extremely poor 75.0mm imaging resolution, the precision is still within 10\% for $ N_{samples} $ = 5000. So, if we can accurately characterize the detector imaging resolution, we can use the $ \chi^2 $ fitting method to adjust for poor imaging performance.
\par
The decay parameter $ \mu = 1/\gamma \beta c \tau_0 $ also plays a significant role in curve fitting. Consider -- as de-excitation lifetime and/or the beam velocity decreases, the source distribution will appear more and more like a point source located at the target (Figure \ref{fig:true_vs_imaged_source_distributions}). This will naturally make it difficult or impossible to identify a de-excitation curve in the image. On the other hand, if the lifetime is too long or the beam velocity too great, then the source distribution will appear as a flat line across the entire imaging range. This also prevents us from reliably fitting a de-excitation curve to the image.
\par
Therefore, in order to image a de-excitation curve we need a sufficient fraction of emissions to occur relatively far from the target while still in the imaging range. As a first-order estimate, for example, we might want a minimum fraction $ F $ of emissions to occur beyond $ z_{min} = 1\sigma_{img} $ of the target, where $ \sigma_{img} $ is our imaging resolution. This sets a lower limit on the lifetime we can measure ($ \tau_{0,min} $). The number of emissions between the target ($ z $ = 0) and $ z_{min} $ is given by:
\begin{align}
\int_{z_{min}}^\infty A_0 e^{-\mu z} dz &= F \int_0^\infty A_0 e^{-\mu z} dz \\
\left( \frac{A_0}{\mu} \right) \left( e^{-\mu z_{min}} \right) &= F \left( \frac{A_0}{\mu} \right) \\
z_{min} &= \frac{\ln{(1/F)}}{\mu} 
\end{align}
Since $ \mu = 1 / (\gamma \beta c \tau) $, the minimum detectable lifetime $ \tau_{0,min} $ is thus:
\begin{equation}
\tau_{0,min} = \frac{z_{min}}{\gamma \beta c \ln{(1/F)}}
\end{equation}
With a beam velocity of 0.30c, imaging resolution of $ \sigma_{img} $ = $ z_{min} $ = 30.0 mm, and $ F $ = 0.25, the minimum detectable lifetime is 230 ps. We can perform a very similar calculation for the opposite extreme -- here we might want some fraction $ F $ of emissions to occur within the imaging window, $ 0 mm < z < 180 mm $. For this case:
\begin{equation}
\tau_{0,max} = \frac{z_{max}}{\gamma \beta c \ln{[1/(1-F)]}}
\end{equation}
For half of all counts to fall between the target and $ z $ = 180 mm ($ F $ = 0.5), the above yields a maximum viable lifetime of 2.7 ns. So, with 30 mm imaging resolution (and unlimited statistics), we would expect to measure lifetimes between 0.23 ns and 2.7 ns.
\par
To study this further, we ran two series of simulations for lifetimes with $ \tau_0 $ varying from 0.029 ns to 6.872 ns. $ \beta $ was kept constant at 0.3c, yielding a $ \mu $ range of 0.365/mm to 0.00154/mm. In the first series, $ N_{samples} $ was kept constant at 5000 counts while we varied imaging resolution $ \sigma_{img} $ from 5.0 mm to 50.0 mm; in the second, $ \sigma_{img} $ was kept constant at 25.0 mm while we varied $ N_{samples} $ from 500 to 50K samples. Each simulation run explored a total of 105 distinct points in the parameter space. (See Tables 2 and 3.)
\par 
Figures \ref{fig:precision_vs_experimental_variables}E-F and \ref{fig:precision_vs_experimental_variables}G-H show a similar nearly monotonically-decreasing systematic error like that in Figures \ref{fig:precision_vs_experimental_variables}C-D. However, here the fits do \textbf{not} converge to the correct lifetime. This is caused by edge effects in the fitting process -- the response functions in the $ P(z,z_0) $ matrix are cut off towards the edges of the imaging window. Rather than having full, symmetric Gaussian shapes at $ z_0 $ = 180 mm, for example, the responses are only half-Gaussians. Because we did not renormalize the detector response functions to account for this potential cut-off, emissions near the edges carry artifically less weight than emissions closer to the target. This causes sharp dips in counts in the idealized lifetime curves near the downstream imaging limit. The effect becomes more pronounced with 1.) long lifetimes, where a greater fraction of emissions occur near the downstream limit; and 2.) very poor imaging resolution, where a larger portion of detector responses get cut off prematurely. It might be possible to improve results by confining curve fitting to a smaller range within the imaging window. One might also extend the imaging window itself and calculate detector responses beyond the $ z $ = {\textpm}180 mm limits used in this study.

\begin{multicolfigure}
\centering
\includegraphics[width=1.0\textwidth]{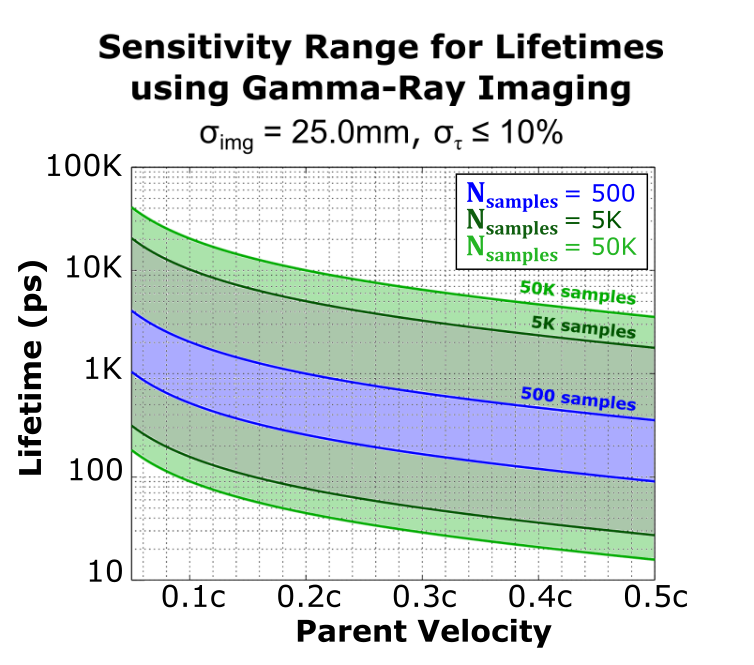}
\captionof{figure}{Sensitive range vs. \# of samples}
$ \sigma_{img} $ = 25.0 mm, 500 to 50K samples \\
Lifetime precision $ \sigma_\tau \leq $ 10\%
\label{fig:lifetime_sensitivity_vs_n_samples}
\end{multicolfigure}

\begin{multicolfigure}
\centering
\includegraphics[width=1.0\textwidth]{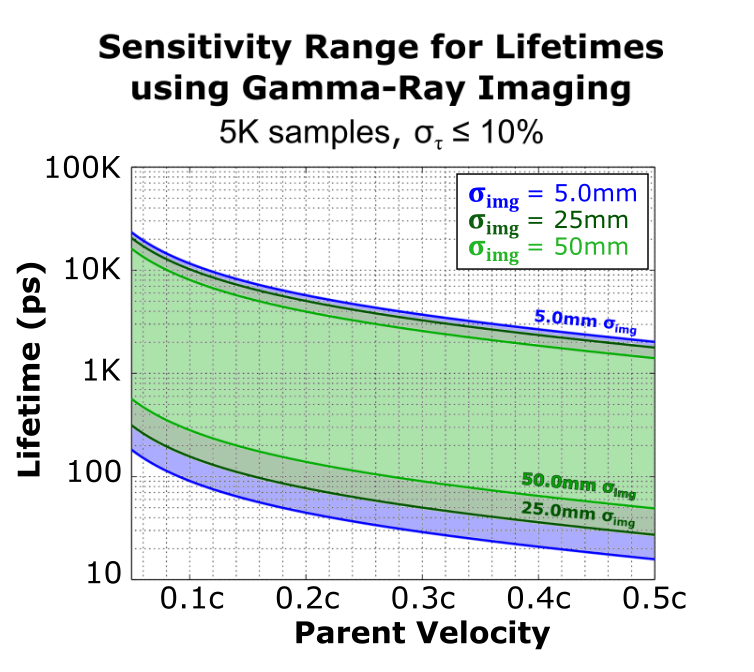}
\captionof{figure}{Sensitive range vs. imaging resolution}
$ \sigma_{img} $ = 5.0 mm to 50.0 mm, 5K samples \\
Lifetime precision $ \sigma_\tau \leq $ 10\%
\label{fig:lifetime_sensitivity_vs_imaging_res}
\end{multicolfigure}

\par
As expected, the uncertainty in the measured $ \tau_0 $ value is most pronounced for very short and very long lifetimes. This suggests an effective range of lifetimes that can be measured through imaging. This sensitivity range will also depend on counting statistics and the uncertainty that can be tolerated in the measured lifetime. The $ \tau_0 $ values in Tables 1 - 3 are marked by relative uncertainty -- with $ \sigma_\tau \leq $10\%, $ \leq $5\%, and $ \leq $1\%. Figures \ref{fig:lifetime_sensitivity_vs_n_samples} and \ref{fig:lifetime_sensitivity_vs_imaging_res} plot the $ \sigma_\tau \leq $ 10\% sensitive ranges as a function of beam velocity from 0.05c to 0.50c. (Again, the decay parameter $ \mu $ is a combined function of $ \beta $ and $ \tau_0 $. As beam velocity increases, the effective range of lifetimes we can measure shifts downward.) The shaded regions in the figures indicate the lifetimes for which we would theoretically measure $ \leq $10\% uncertainty, given a specific imaging resolution and number of counts. Note that some of the ranges are cut off. For example, with 5K counts and 10.0mm imaging resolution, the observed uncertainty in measured lifetime is less than 10\% even at 29 ps -- the shortest lifetime for which we ran simulations.

%%%%%%%%%%%%%%%%%%%%
% Section: Experimental Validation
%%%%%%%%%%%%%%%%%%%%
\section{Experimental Validation}
\label{sec:experimental_validation}

We chose two known transitions in \textsuperscript{92}Mo to validate the lifetime measurement techniques described above. The experiment was performed at ANL's ATLAS facility, using a \textsuperscript{84}Kr projectile on a \textsuperscript{12}C target to populate the desired nuclear states through the \textsuperscript{12}C[\textsuperscript{84}Kr, 4n]\textsuperscript{92}Mo reaction. The \textsuperscript{84}Kr projectiles were accelerated to 394 MeV \textit{total} kinetic energy, resulting in a pre-collision speed of 0.100c. The \textsuperscript{92}Mo nuclei thus produced had a maximum recoil speed of 0.085c. The experimental detector geometry is shown in Figure \ref{fig:detector_geometry_anl}; it gives the angular distribution of GRETINA detector modules as seen from the beam target at (0, 0, 0). The dark blue points show where photon hits registered in each detector crystal, giving an idea of the angular distribution of emissions. \cite{anl_mo92_presentation}

\begin{multicolfigure}
\centering
\includegraphics[width=1.0\textwidth]{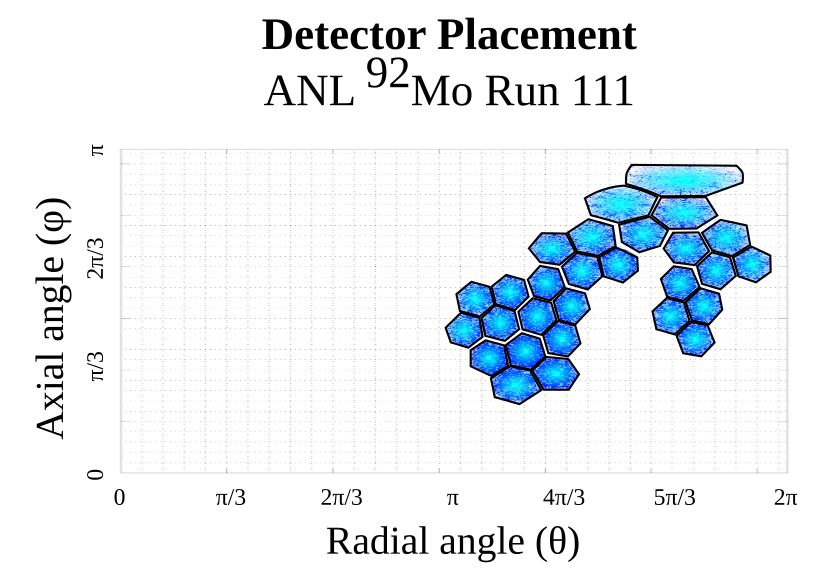}
\captionof{figure}{Angular distribution of detector}
modules as seen from the beam target
\label{fig:detector_geometry_anl}
\end{multicolfigure}

\par
We used a Doppler-corrected emission spectrum to find suitable transitions from the excited \textsuperscript{92}Mo nuclei. Compton sequencing was used to identfiy the first hit in each detected photon track. By assuming all detected gamma rays were emitted from the beam target, we obtained a corresponding estimate for each emission angle. This let us convert the detected lab-frame energies into a CM-frame energy spectrum (shown in Figure \ref{fig:corrected_mo92_spectrum}). A 2064.1 keV line is plainly visible along with many lower-energy lines.

\begin{multicolfigure}
\centering
\includegraphics[width=1.0\textwidth]{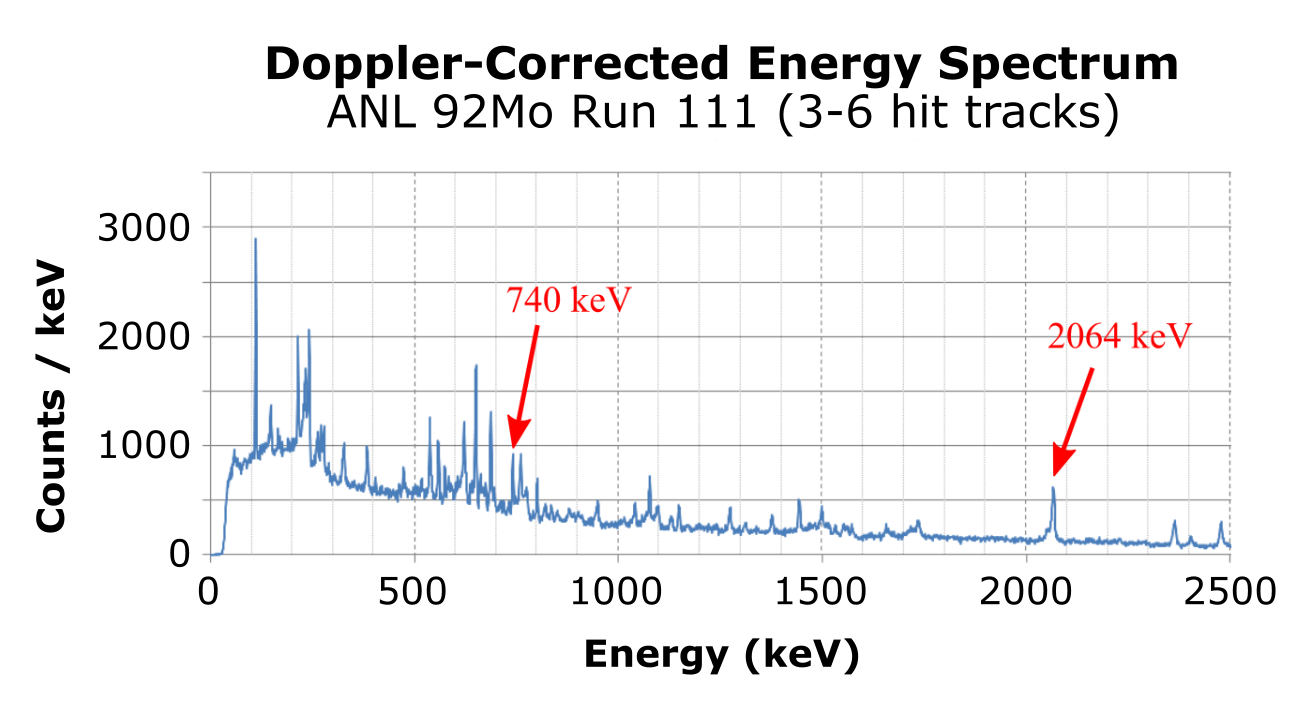}
\captionof{figure}{Doppler-Corrected \textsuperscript{92}Mo Spectrum}
\label{fig:corrected_mo92_spectrum}
\end{multicolfigure}

\par
We had two primary requirements for the transition used in the lifetime measurement -- a strong gamma peak and a long enough lifetime. 
The emitted gamma ray would need to stand out of the energy spectrum, not overlapping with nearby peaks; and the source distribution would need to be broader than our expected imaging resolution. Figure \ref{fig:mo92_gamma_lines} lists several candidates taken from Brookhaven National Laboratory's Chart of Nuclides database. \cite{nndc_chart_of_nuclides} The table includes the energy windows used to define each peak, as well as peak-to-totals and signal-to-noise ratios for those windows. 

\begin{multicolfigure}
\centering
\includegraphics[width=1.0\textwidth]{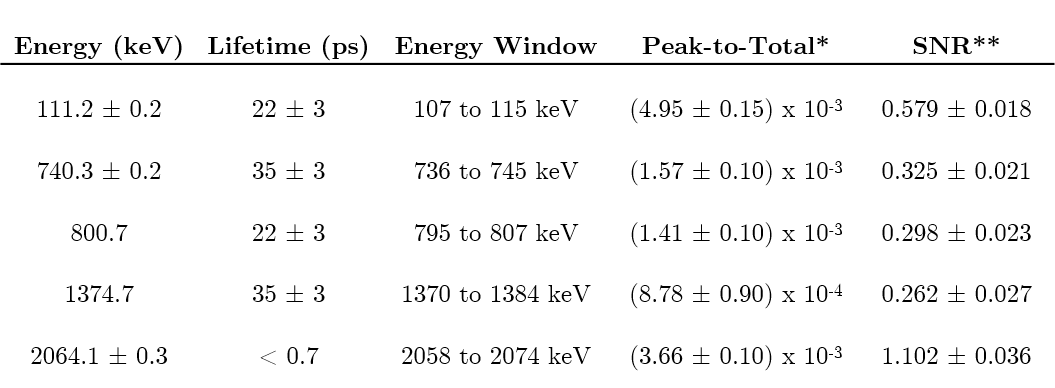}
\begin{footnotesize}
* P/T = Peak net counts to \textbf{total} counts (0-3000 keV) \\
** SNR = Peak net counts to peak background counts \\
\end{footnotesize}
\captionof{figure}{Common $ \gamma $-ray lines of \textsuperscript{92}Mo}
\label{fig:mo92_gamma_lines}
\end{multicolfigure}

\par
We chose to use the 740.3 keV and 2064.1 keV transitions. While the 111.2 keV peak is the strongest one in the spectrum, it is too low-energy for Compton imaging. The state's lifetime is also on the extreme low-end of what we expect to be able to see with either imaging mode. (The decay parameter of $ \mu $ = 1.8 mm\textsuperscript{-1} means that 99.9\% of the nuclei will de-excite within 3.9 mm of the target.) The 800.7 keV peak was discarded for the same reason. Meanwhile, the 740.3 keV and 1347.7 keV transitions are part of a de-excitation chain and share a lifetime on the low end of the sensitive range we found in our simulations. We chose the 740.3 keV line because its peak-to-total was significantly higher.

\begin{multicolfigure}
\centering
\includegraphics[width=0.85\textwidth]{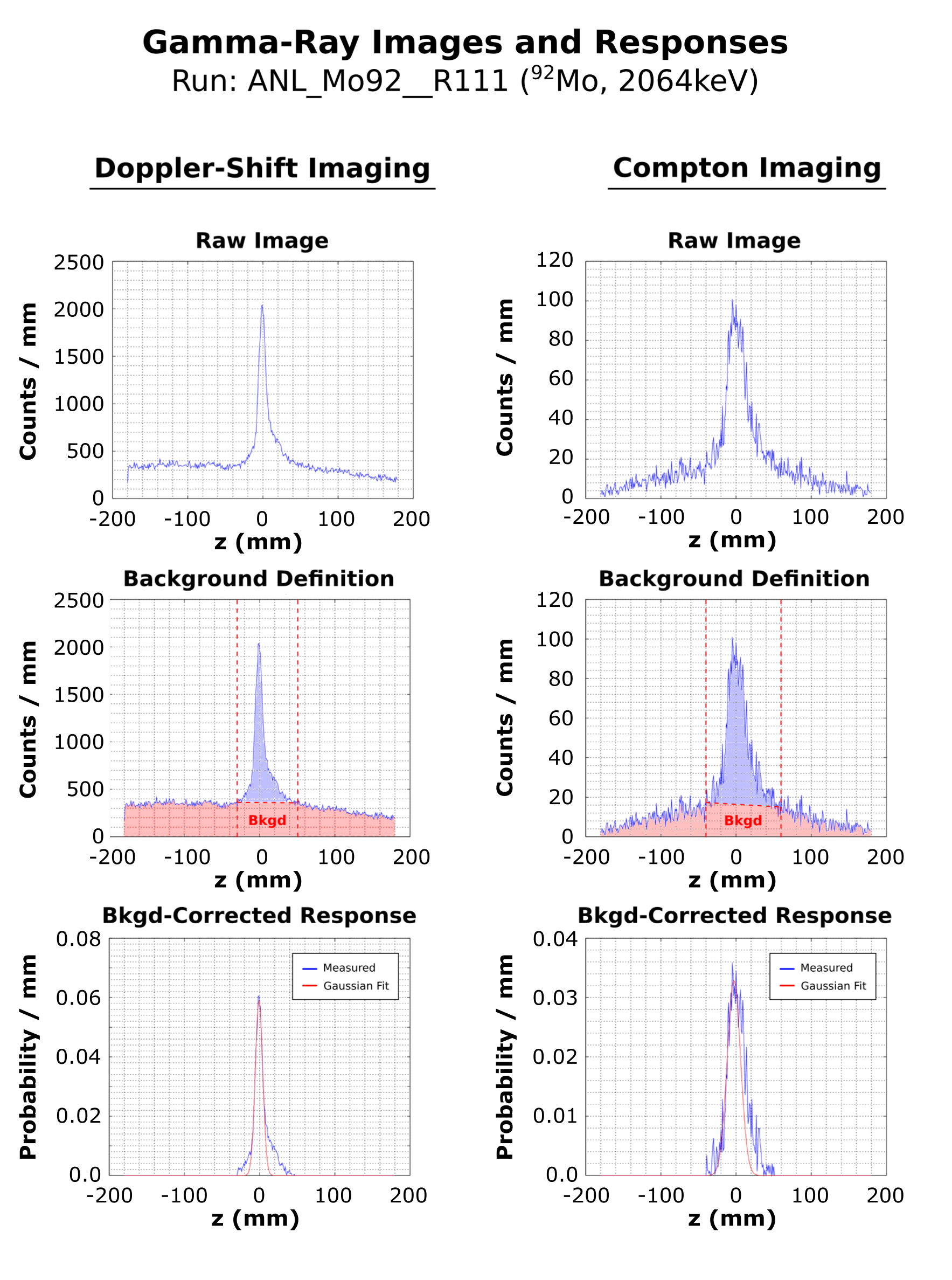}
\captionof{figure}{Imaging responses at 2064.1 keV}
\label{fig:imaging_responses__anl_mo92__r111__2064keV}
\end{multicolfigure}

\begin{multicolfigure}
\centering
\includegraphics[width=0.85\textwidth]{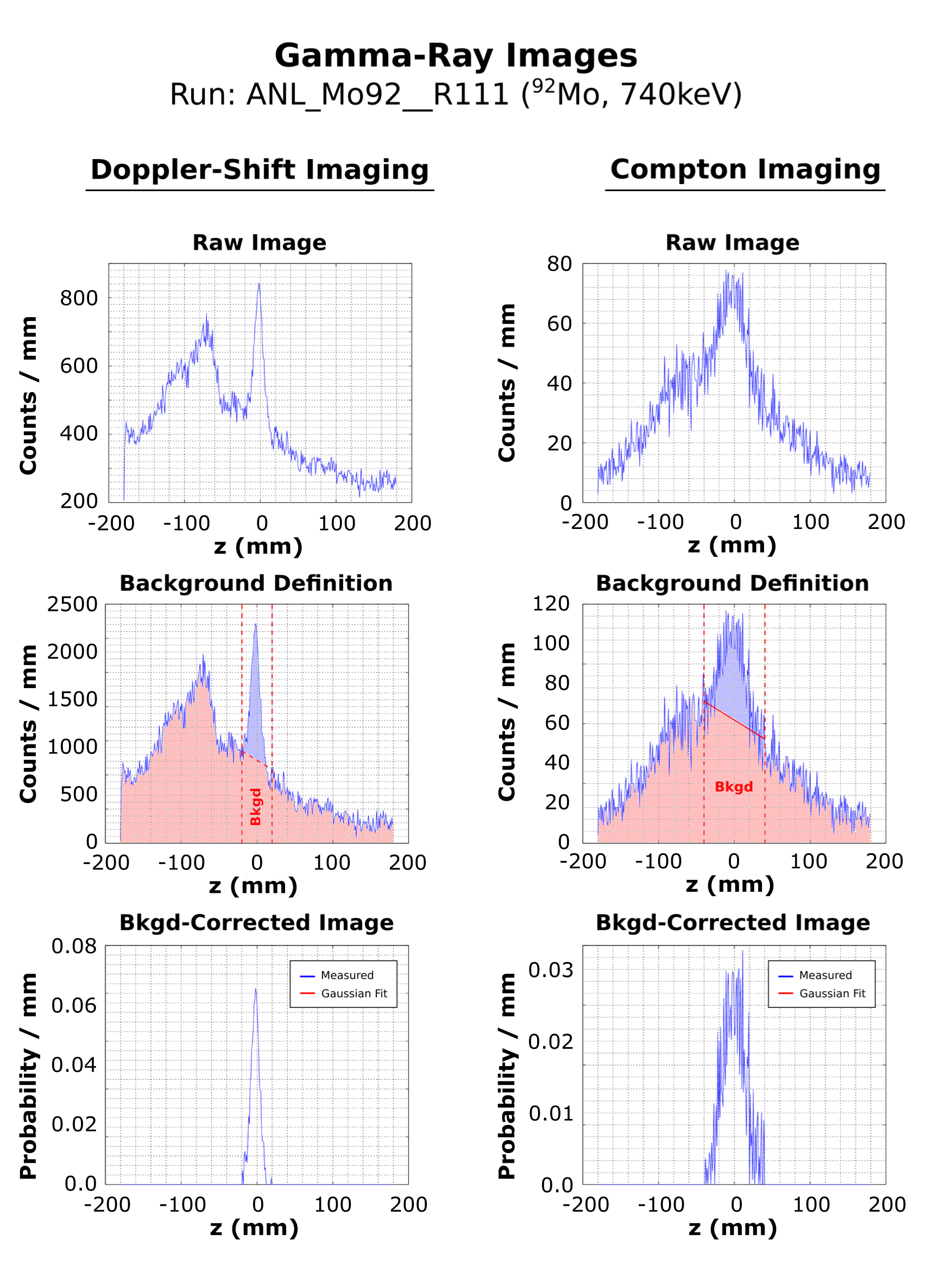}
\captionof{figure}{$ \gamma $-ray images at 740.3 keV}
\label{fig:imaging_responses__anl_mo92__r111__740keV}
\end{multicolfigure}

\par
The 2064.1 keV transition is useful specifically because of its \textit{very} short known lifetime of $ < $ 0.7 ps. Combined with the low 0.085c parent velocity, this yields a decay parameter of $ \mu > $ 56 mm\textsuperscript{-1}. Neither gamma-ray imaging method has sufficiently good resolution to distinguish such a short de-excitation curve -- 99.9\% of the nuclei de-excite within the first 0.12 mm downstream of the target. This is actually an advantage in this context, because it means the emissions can be approximated as a point source at the target ($ z_0 $ = 0.0 mm). With the true emission point known, we can therefore treat the gamma-ray \textit{image} as the imaging \textit{response}.
\par
We used the imaging response measured for 2064.1 keV to produce a $ \chi^2 $ fit of the image produced for 740.3 keV. Doing so assumes that the imaging responses are similar at both energies, although this is only an approximation. (See Figures \ref{fig:imaging_responses__anl_mo92__r111__2064keV} and \ref{fig:imaging_responses__anl_mo92__r111__740keV}.) At 2064.1 keV, both Compton and Doppler-Shift imaging produce single-peak images, but they have somewhat different shapes than those produced in simulations of a 1.0 MeV beam source. There is also noticeably higher background. The Doppler-shift image, in particular, resembles a Gaussian sitting atop a nearly uniform background. Unlike the simulations, the energy spectrum in this experiment continues past the photopeak energy. Recall that lower-energy background below a peak causes a low tail of counts on the right side of the image. Similarly, higher-energy background will give rise to a low tail on the \textit{left} half of the image.
\par
The images at 740.3 keV are less well-behaved. The Doppler-shift image still exhibits a clear peak near the target; this reflects the sharp de-excitation curve for the 740.3 keV transition. However, the background is far from uniform across the imaging range. In fact, there appears to be a second, broader peak to the left of the target. This is due to another nearby photopeak at 762.9 keV, which is too close in energy to be rejected directly. The Compton image is also degraded compared to the corresponding image at 2064.1 keV. From Figures \ref{fig:corrected_mo92_spectrum} and \ref{fig:mo92_gamma_lines}, we can see that the Compton background at 740.3 keV is significantly higher than at 2064.1 keV. While many of these background counts can be rejected by data filters, the ones that remain will blur the output images.
\par
To apply the detector response from 2064.1 keV to a lifetime measurement for the 740.3 keV transition, we need a way to reduce both imaging responses to comparable forms. Figures \ref{fig:imaging_responses__anl_mo92__r111__2064keV} and \ref{fig:imaging_responses__anl_mo92__r111__740keV} illustrate our approach. First, we obtain a raw gamma-ray image (top row). Next, average background is estimated by defining a window about the observed ``peak'' (middle row). Requiring counts to go to zero at window boundaries produces a trapezoidal approximation of the background within the peak area. Lastly, this background is subtracted from the total counts to isolate the ``true'' image (bottom row). In the case of the 2064.1 keV transition, this became the imaging response used in the lifetime measurement.
\par
This fairly crude approach could be improved. Imaging response depends on more than a photon's CM-frame energy or the properties of the parent nuclei. It also depends on the highly complex structure of the background in the allowed energy range (in this case, roughly 680 - 795 keV). This makes it difficult to use imaging responses at other energies or from other nuclei as surrogates in lifetime measurements. It is also unclear how one would use this method for de-excitation curves that span a greater portion of the imaging window (i.e. due to longer lifetime and/or higher parent velocity). As the 740.3 keV image shows, we might not be able to use a trapezoidal estimate for background in such cases.
\par
Figure \ref{fig:measured_mo92_data} shows the results of the experimental measurements. We restricted tracks to 3-6 hits in length and the CM-frame energies to the ranges in Figure \ref{fig:mo92_gamma_lines}, but otherwise did not apply any data cuts. Gaussian imaging resolution was only calculated for the 2064.1 keV transition, which -- due to its very short lifetime -- could be treated as a point source. Imaging efficiencies were taken as the ratio of successful reconstructions to the number of counts in their respective peak energy windows. We found that Doppler-shift imaging yielded close to 5 times better resolution and 10 times better imaging efficiency than Compton imaging in this test. 

\begin{multicolfigure}
\centering
\includegraphics[width=1.0\textwidth]{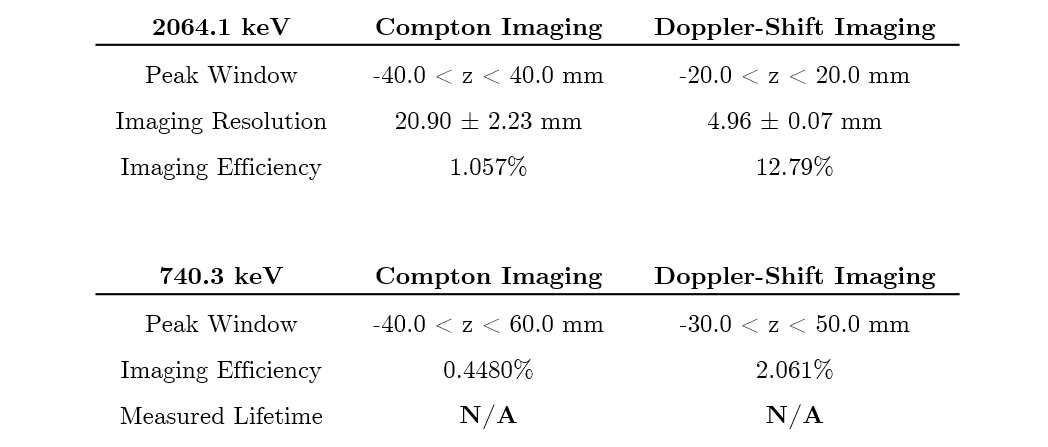}
\captionof{figure}{Lifetime measurement for \textsuperscript{92}Mo}
\label{fig:measured_mo92_data}
\end{multicolfigure}

\par
Unfortunately, we were unable to obtain a non-zero lifetime for the 704.3 keV \textsuperscript{92}Mo transition. With both Compton and Doppler-shift imaging, the 2064.1 keV imaging response was wider than the image produced by the 740.3 keV line -- in other words, imaging resolution was insufficient to resolve the de-excitation curve. Because of this, the $ \chi^2 $ algorithm described in Section \ref{sec:curve_fitting_and_imaging} converged to a zero lifetime.
\par
A primary limitation here is the combination of short nuclear lifetime (35 ps) \& low recoil speed (0.085c). The decay parameter of $ \mu = 1 / \gamma \beta c \tau_0 $ = 1.1 mm\textsuperscript{-1} means that 99.9\% of the nuclei will de-excite within 6.2 mm of the target. Raising the energy of the \textsuperscript{84}Kr projectile would result in a faster recoil speed for the \textsuperscript{92}Mo nuclei, and might improve results. For example, increasing speed to 0.3c would result in a decay parameter of $ \mu $ = 0.30 mm\textsuperscript{-1}, giving a mean flight distance of 3.3 mm and nearly 4 times more distance over which to resolve a de-excitation curve. The 4.96 mm imaging resolution obtained with Doppler-shift imaging would be sufficient for the source to no longer appear point-like.
\par
However, countrate might also become a factor, and may necessitate longer beam runtimes. Increasing beam energy often comes at the cost of beam current. Nuclear reaction cross-sections may also be significantly affected depending on the projectile and target used. We had sufficient counts in this experiment -- the 2064.1 keV peak was characterized with about 3.5K and 35K counts for Compton and Doppler-shift imaging, respectively. Meanwhile, the 740.3 keV peak had about 1.4K counts for Compton imaging and 5.8K for Doppler-shift imaging, respectively. This does not give very much room for reducing total counts. As shown in Table 1, we expect to need $ > $500 counts to achieve $ < $10\% relative uncertainty even with a fairly ideal lifetime of 859 ps.

%%%%%%%%%%%%%%%%%%%%
% Section: Future Work
%%%%%%%%%%%%%%%%%%%%
\section{Future Work}
\label{sec:future_work}

More work needs to be done to refine these measurement techniques. One problem lies in accurately characterizing the detector response for the transition under study. There are a few methods one might try to improve this characterization. For example, Compton imaging is able to image stationary sources. We can therefore place a calibration point source (like \textsuperscript{60}Co) at a fixed location on the beam axis, and use the measured image as the detector response. This method does assume, however, that the velocity of the beam has little effect on the overall imaging response. This method therefore does \textit{not} work for Doppler-shift imaging.
\par
We could also make use of the $ \chi^2 $ fitting technique used in lifetime measurements (Section \ref{sec:curve_fitting_and_imaging}). Instead of starting with a \textit{known} detector response and an unknown lifetime, we could use a transition with known \textit{lifetime} to deduce the unknown detector response. (The 2064.1 keV transition of \textsuperscript{92}Mo is a good example.) The goal would be to find a hypothesized response that most closely transforms the known source distribution (determined by lifetime and parent velocity) into the measured image. One might experiment with different \textit{shapes} and parameterizations for the response, and decide whether to account for geometric complications such as detector placement. With a properly-parametrized detector response, one could construct a $ \chi^2 $ curve to find the best-fit value for each parameter. 
\par
However, it may not be suitable to use the detector response from one transition as a surrogate response for a different energy. One potential solution is a sufficiently-realistic simulation. We have only explored a simple 1.0 MeV beam source, but one could simulate other experimental setups with different detector resolution and nuclear properties. The goal would be to run such simulations for a range of lifetimes, producing a different response matrix for each, and then find the one that produced the best experimental results. One could potentially quantify this by looking for the response that yielded the smallest uncertainty on the measured lifetime. 
\par
Despite the negative results of this experimental validation, imaging may still have use in lifetime measurements. The 740.3 keV transition of \textsuperscript{92}Mo has a very short lifetime. Combined with the low recoil velocity of the parent nuclei, this de-excitation was on the low-end of the expected sensitive range. It's likely there are other states that would be more conducive to study with gamma-ray imaging.

%%%%%%%%%%%%%%%%%%%%%%%%%%%%%%%%%%%%%%%%
% Section: Bibliography
%%%%%%%%%%%%%%%%%%%%%%%%%%%%%%%%%%%%%%%%

%%%%%%%%%%%%%%%%%%%%%%%%%%%%%%%%%%%%%%%%
% End-of-Document Figures & Tables
%%%%%%%%%%%%%%%%%%%%%%%%%%%%%%%%%%%%%%%%

\end{multicols}

\begin{center}
\centering 
~\\
~\\
{\LaTeX} is a relic of the 1970's that suffers from poor UX, uninformative error messages, and \\
unpredictable output. MS Word \& WYSIWYG will make you \& your team more productive. \\
Would you rather spend your time typesetting, or researching?
~\\
~\\
\textit{An Efficiency Comparison of Document Preparation Systems} \\
\textit{Used in Academic Research \& Development} \\
Markus Knauff, Jelica Nejasmic \\
PLoS One 9(12): e 115069. doi:10.1371/journal.pone.0115069
\end{center}

% Main figures for precision in lifetime measurements

% Raw data
\includepdf{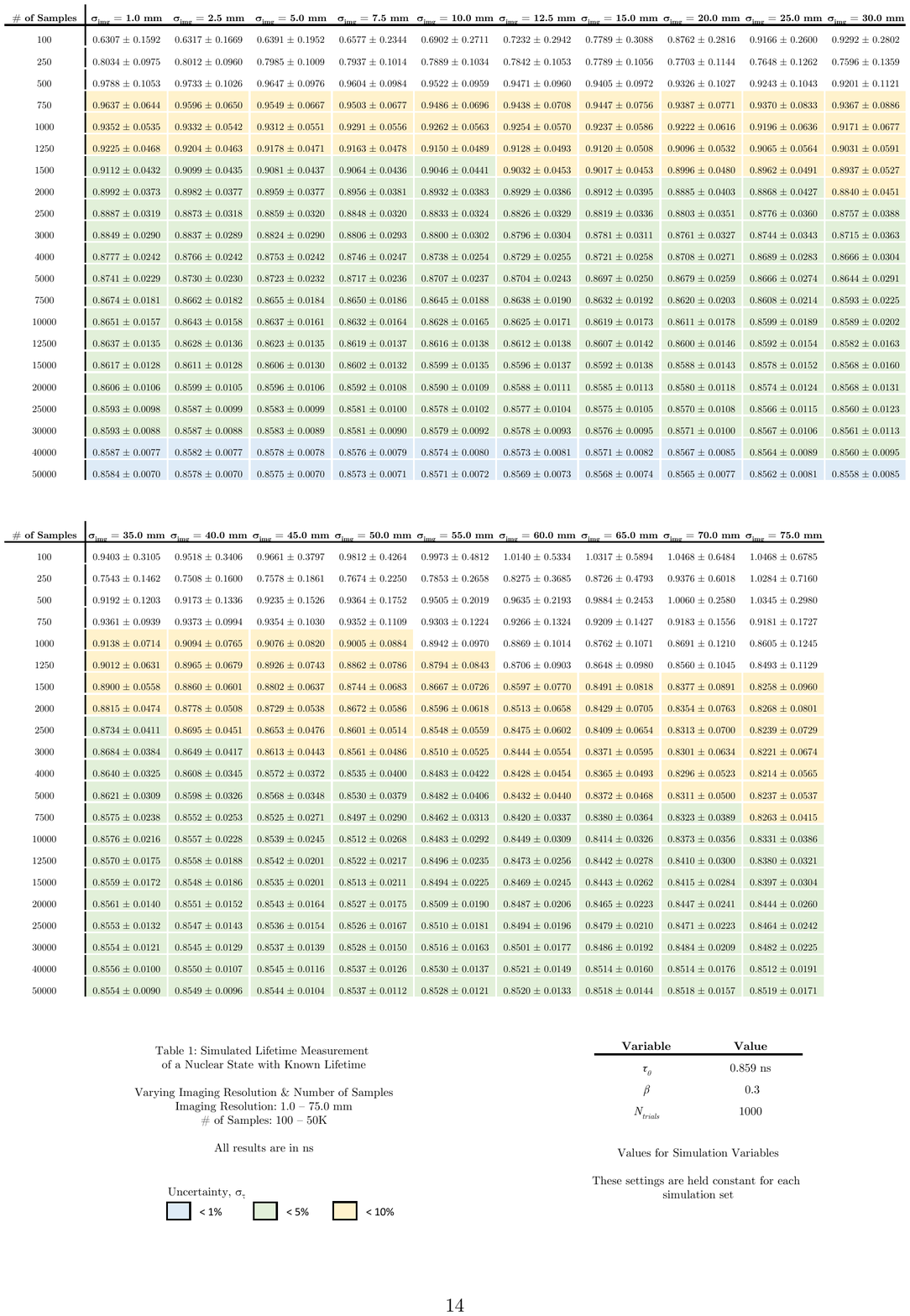}
\includepdf{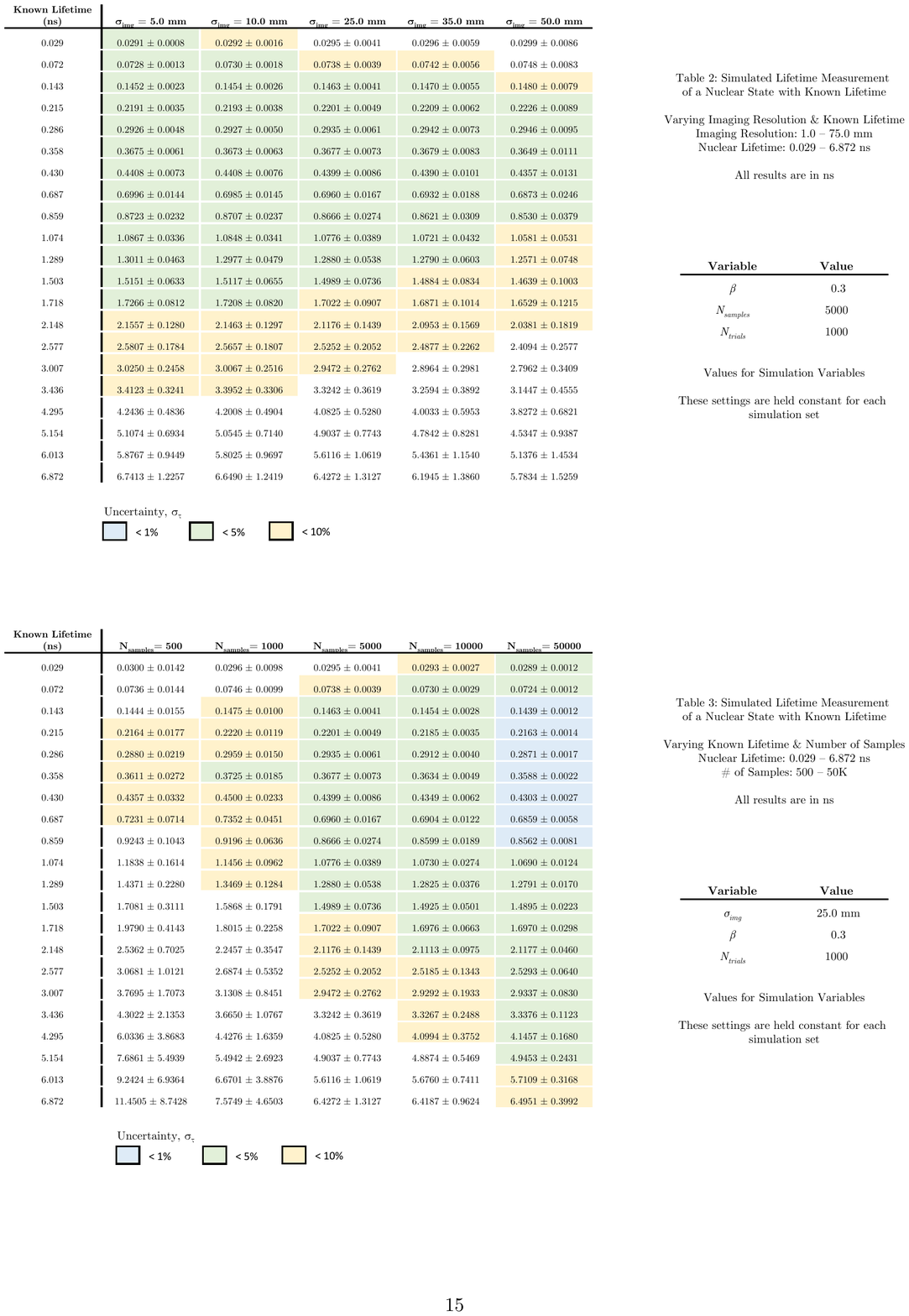}

%\par
%\incomplete{CURRENT PROGRESS}


\begin{thebibliography}{10}

% I-Yang's paper on gamma-ray tracking
\bibitem{gamma_ray_tracking_detectors}
I.Y. Lee.
\textit{Gamma-Ray Tracking Detectors}.
1999.
Nuclear Instrumentation and Methods in Physics Research A, Vol. 422 Pps. 195-200.
\url{http://www.sciencedirect.com/science/article/pii/S0168900298010936}

% Main GRETA paper about using Compton tracking
\bibitem{new_concepts_in_gamma_detection}
M.A. Deleplanque, I.Y. Lee, K. Vetter, et al.
\textit{GRETA: Utilizing New Concepts in Gamma-ray Detection}
1999.
Nuclear Instruments and Methods in Physics Research A, Vol. 430 Pps. 292-310.
\url{http://www.sciencedirect.com/science/article/pii/S0168900299001874}

% NSAC 2015 roadmap 
\bibitem{nsac_long_range_plan}
DOE NSAC
\textit{Reaching for the Horizon: The 2015 Long Range Plan for Nuclear Science}. 
\url{https://science.osti.gov/-/media/np/nsac/pdf/2015LRP/2015_LRPNS_091815.pdf}

% NuPECC 2017 roadmap 
\bibitem{nupecc_long_range_plan}
NuPECC
\textit{Long Range Plan 2017: Perspectives in Nuclear Physics}. 
\url{http://www.nupecc.org/lrp2016/Documents/lrp2017.pdf}

% Pitch for AGATA and GRETA for gamma-ray spectroscopy
\bibitem{agata_and_greta}
I-Yang Lee and J. Simpson.
\textit{AGATA and GRETA: The Future of Gamma-Ray Spectroscopy}.
2010.
Nuclear Physics News, Vol. 20 Pps. 23-28. 
\url{https://doi.org/10.1080/10619127.2010.506124}.

% Initial performance of GRETINA for spectroscopy
\bibitem{gretina_performance_whitepaper}
S. Paschalis, I.Y. Lee, A.O. Macchiavelli, et al.
\textit{The Performance of the Gamma-Ray Energy Tracking In-beam Nuclear Array GRETINA}.
2013.
Nuclear Instruments and Methods in Physics Research A, Vol. 709 Pps. 44-55. 
\url{http://www.sciencedirect.com/science/article/pii/S0168900213000508}

% Performance of GRETINA for spectroscopy at NSCL
\bibitem{gretina_spectroscopy_performance}
D. Weisshaar, D. Bazin, P. Bender, et al.
\textit{The Performance of the Gamma-Ray Tracking Array GRETINA for Gamma-Ray Spectroscopy with Fast Beams of Rare Isotopes}
2017.
Nuclear Instruments and Methods in Physics Research A, Vol. 847 Pps. 187-198. 
\url{http://www.sciencedirect.com/science/article/pii/S0168900216312402}.

% GRETA homepage
\bibitem{greta_website}
GRETA Collaboration.
\textit{GRETA: The Gamma-Ray Energy Tracking Array}. 
2021.
\url{https://sites.google.com/a/lbl.gov/greta/}.

% Main AGATA whitepaper
\bibitem{agata_whitepaper}
S. Akkoyun, A. Algora, B. Alikhani, et al.
\textit{AGATA: Advanced GAmma Tracking Array}.
2012.
Nuclear Instruments and Methods in Physics Research A, Vol. 668 Pps. 26-58. 
\url{http://www.sciencedirect.com/science/article/pii/S0168900211021516}.

% AGATA homepage
\bibitem{agata_website}
AGATA Collaboration.
\textit{AGATA: The Advanced GAmma Tracking Array}.
2021.
\url{https://www.agata.org/}.

% Brought up in Ch.2 when discussing tech specs for GRETINA array
%\bibitem{GRETINA_tech_specs}
%GRETINA.
%\textit{GRETINA Project Development}.
%\url{http://gretina.lbl.gov/project-development}.
%Published 2016-05-10.

% Goes into many details about GRETA. I took a schematic from it, too
%\bibitem{GRETA_whitepaper}
%GRETINA Users Executive Committee.
%\textit{The Gamma-Ray Energy Tracking Array Whitepaper}.
%Nuclear Astrophysics and Low Energy Nuclear Physics Town Meeting, Texas A\&M.
%2014.

% Kai's paper on the potential of gamma-ray tracking
\bibitem{gamma_ray_tracking_opportunities}
K. Vetter.
\textit{Gamma-Ray Tracking: New Opportunities for Nuclear Physics}.
2002.
Nuclear Physics News, Vol. 12 \#2, P. 15-20.

% Simulation groundwork for Compton imaging in GRETINA
\bibitem{compton_imaging_in_gretina}
R. Crabbs and I. Lee and K. Vetter.
\textit{Using Compton Imaging to Locate Moving Gamma-Ray Sources in the GRETINA Detector}.
\url{arXiv.org}.
Published 2020, Accessed 2020-07-05.

% Simulation groundwork for Doppler-Shift imaging in GRETINA
\bibitem{doppler_imaging_in_gretina}
R. Crabbs and I. Lee and K. Vetter.
\textit{Using Doppler-Shift Imaging to Locate Moving Gamma-Ray Sources in the GRETINA Detector}.
\url{arXiv.org}.
Published 2020, Accessed 2020-07-05.

% Reference for RDM concept diagram
%\bibitem{rdm_lifetime_measurements}
%Isotope Science Facility at Michigan State University.
%\textit{Upgrade of the NSCL Rare Isotope Research Capabilities}.
%2006.
%MSUCL Vol. 1345.

% An example lifetime measurement with GRETINA
\bibitem{lifetime_measurement_74rb}
C. Morse, H. Iwasaki, A. Lemasson, et al.
\textit{Lifetime Measurement of the 2+1 state in \textsuperscript{74}Rb and Isospin Properties of Quadrupole Transition Strengths at N = Z}.
2018.
Physics Letters B, Vol. 787 Pps. 198-203.
\url{https://www.sciencedirect.com/science/article/pii/S037026931830844X}.

% Description of TRIPLEX plunger lifetime measurements
\bibitem{rdm_triplex_plunger}
H. Iwasaki, A. Dewald, T. Braunroth, et al.
\textit{The TRIple PLunger for EXotic beams: TRIPLEX for Excited-State Lifetime Measurement Studies on Rare Isotopes}.
2016.
Nuclear Instruments and Methods in Physics Research A, Vol. 806, 123-131, doi; http://dx.doi.org/10.1016/j.nima.2015.09.091. URL 
\url{https://www.sciencedirect.com/science/article/pii/S0168900215011626}.

% Results for Compton sequencing in GRETINA 
\bibitem{compton_sequencing_in_gretina}
R. Crabbs and I. Lee and K. Vetter.
\textit{Simulations of Compton Sequencing with the GRETINA Detector}.
\url{arXiv.org}.
Published 2020, Accessed 2020-12-18.

% Krane textbook, pages 170-172, 173 (Bateman Equations)
%\bibitem{krane_textbook}
%K.S. Krane.
%\textit{Introductory Nuclear Physics, 3rd Edition}.
%Wiley. Published 1987.
%Pages 170-173.

% Kai's paper on position resolution in segmented detectors
\bibitem{position_sensitivity_in_hpge}
K. Vetter and others.
\textit{Three-Dimensional Position Sensitivity in Two-Dimensionally Segmented HPGe Detectors}.
2000.
Nuclear Instrumentation \& Methods in Physics Research A, Vol. 452 P. 223-238.

% Kai's paper on GRETA detector performance
\bibitem{greta_detector_performance}
K. Vetter and others.
\textit{Performance of the GRETA Prototype Detectors}.
2000.
Nuclear Instrumentation \& Methods in Physics Research A, Vol. 452 P. 105-114.

% Series of talks with Chris and Lew about the UCGretina Geant4 model
\bibitem{UCGretina_model}
L. Riley and C. Campbell and H. Crawford.
\textit{UCGretina v3.0 Geant4 model}.
2015. 
Personal communication.

% Description of Chi2 fitting method
\bibitem{chi2_fitting_method}
Wolfram Alpha.
\textit{Fitting Data to Linear Models by Least-Squares Techniques}.
\url{http://reference.wolfram.com/applications/eda/FittingDataToLinearModelsByLeast-SquaresTechniques.html}.
Accessed 2016-05-11.

% Reference for how SVD works
\bibitem{SVD_fitting_method}
E. Weisstein.
\textit{Singular Value Decomposition}.
Wolfram Mathworld.
\url{http://mathworld.wolfram.com/SingularValueDecomposition.html}.
Accessed 2016-07-21.

% Reference for the noise amplification problem in SVD
\bibitem{noise_in_SVD}
Y. Shim and Z. Cho.
\textit{SVD Pseudoinversion Image Reconstruction}.
IEEE Transactions on Acoustics, Speech, and Signal Processing.
Vol. 29 \#4, Pages 904-909.
Published 2003.

% Mentioned briefly when talking about SVD and choice of binning
%\bibitem{ideal_binning}
%A.J. Izenman.
%\textit{Recent Developments in Nonparametric Density Estimation}.
%Journal of the American Statistical Association.
%Vol. 86 \#413, Pages 205-224. 
%Published 1991.

% Powerpoint presentation describing Mo-92 experiments at ANL
\bibitem{anl_mo92_presentation}
T. Lauritsen.
\textit{Shootout Experiment GSFMA315 at a Glance}.
Argonne National Laboratory.
Presentation, 2014-04-30.

% NNDC Chart of Nuclides, used for the Mo-92 experimental validation
\bibitem{nndc_chart_of_nuclides}
Brookhaven National Laboratory.
\textit{Chart of Nuclides}.
\url{http://www.nndc.bnl.gov/chart/}.
Accessed 2016-06-20.

\end{thebibliography}
\end{document}